\documentclass{aa}
\usepackage{xcolor}

\definecolor{Ca4}{HTML}{FCAF93}
\definecolor{Ca3}{HTML}{F44F39}
\definecolor{Ca2}{HTML}{AA1016}
\definecolor{Sa4}{HTML}{BAD6EB}
\definecolor{Sa3}{HTML}{539ECD}
\definecolor{Sa2}{HTML}{0B559F}
\definecolor{Ga4}{HTML}{56B567}

\usepackage{natbib}
\usepackage{graphicx}
\usepackage{stfloats}
\usepackage{placeins}
\usepackage[flushleft]{threeparttable}
\usepackage{txfonts}
\usepackage[breaklinks,colorlinks,urlcolor=blue,citecolor=blue,linkcolor=blue]{hyperref}
\begin{document}

   \title{Constraining giant planet formation with synthetic ALMA images of the Solar System's natal protoplanetary disk}

   \author{C. Bergez-Casalou\inst{1}, B. Bitsch\inst{1}, N.T. Kurtovic\inst{1}, P. Pinilla\inst{1,}\inst{2}}

   \institute{\inst{1}Max-Planck-Institut für Astronomie, Königstuhl 17, 69117 Heidelberg, Germany, email:bergez@mpia.de\\
              \inst{2}Mullard Space Science Laboratory, University College London, Holmbury St Mary, Dorking, Surrey RH5 6NT, UK}


 
 \abstract
  {New ALMA observations of protoplanetary disks allow us to probe planet formation in other planetary systems, giving us new constraints on planet formation processes. Meanwhile, studies of our own Solar System rely on constraints derived in a completely different way. However, it is still unclear what features the Solar System protoplanetary disk could have produced during its gas phase. By running 2D isothermal hydro-simulations used as inputs for a dust evolution model, we derive synthetic images at millimeter wavelengths using the radiative transfer code RADMC3D. We find that the embedded multiple giant planets strongly perturb the radial gas velocities of the disk. These velocity perturbations create traffic jams in the dust, producing over-densities different from the ones created by pressure traps and located away from the planets' positions in the disk. By deriving the images at $\lambda =\rm 1.3mm$ from these dust distributions, we show that very high resolution observations are needed to distinguish the most important features expected in the inner part ($< 15 $AU) of the disk. The traffic jams, observable with a high resolution, further blur the link between the number of gaps and rings in disks and the number of embedded planets. We additionally show that a system capable of producing eccentric planets by scattering events that match the eccentricity distributions in observed exoplanets does not automatically produce bright outer rings at large radii in the disk. This means that high resolution observations of disks of various sizes are needed to distinguish between different giant planet formation scenarios during the disk phase, where the giants form either in the outer regions of the disks or in the inner regions. In the second scenario, the disks do not present planet-related features at large radii. Finally, we find that, even when the dust temperature is determined self-consistently, the dust masses derived observationally might be off by up to a factor of ten compared to the dust contained in our simulations due to the creation of optically thick regions. Our study clearly shows that in addition to the constraints from exoplanets and the Solar System, ALMA has the power to constrain different stages of planet formation already during the first few million years, which corresponds to the gas disk phase.} 


   \authorrunning{C.Bergez-Casalou et al}
   \titlerunning{Constraining planet formation during the disk phase}

   \maketitle
%

\section{Introduction}

Recent observations with the Atacama Large Millimeter/Submillimeter Array (ALMA) and  the Spectro-Polarimetric High-contrast Exoplanet REsearch (SPHERE) instruments show protoplanetary disks that present different kinds of substructures (rings, gaps, cavities, and asymmetries) present in the gas \citep[e.g.,][]{Teague2018,Pinte2020} and in the dust \citep[e.g.,][]{ALMA2015,Avenhaus2018,Andrews2018}. These substructures may have different possible origins, including: self-induced dust traps due to dust growth and dust backreaction on the gas \citep{Gonzalez2017}, dust growth in snow lines \citep{Zhang2015}, zonal flows \citep{Flock2015}, secular gravitational instabilities \citep{Takahashi2016,Tominaga2020}, sintering-induced rings \citep{Okuzumi2016}, and gap opening embedded planets \citep{Pinilla2012}. 

Focusing on the features created by planets, it is hard to observe the planets directly while they are embedded in their protoplanetary disk \citep{Sanchis2020,Kloster2021,AsensioTorres2021}. Therefore, analyzing the dust gap size \citep{Zhang2018} or the CO velocity perturbations \citep{Teague2018,Pinte2020} are ways to indirectly derive the properties of potentially embedded planets. Assuming that these features are indeed caused by planets, we are able to probe forming planets that are not observable directly and that will continue to evolve by accretion and migration in disks. These objects can then be used to derive or confirm some constraints on planet formation processes.

On the other hand, our own Solar System has some characteristics representative of its birth environment. For example, meteorites are solid remains of the protoplanetary solid disk. \cite{Kruijer2017} showed that their chemical composition in the Solar System can be used to constrain the time at which Jupiter's core formed, as it is supposed to separate the reservoirs of carbonaceous and non-carbonaceous chondrites by blocking the pebbles flowing through the disk. Using solid mass estimates from the asteroid and Kuiper belts, \cite{Lenz2020} tried to reproduce the possible gas and solid distributions of our natal protoplanetary disk. 

Different models investigate how different parts of the Solar System could have formed. For example, the classical model \citep{Wetherill1994,Raymond2009b} attempts to reproduce the inner Solar System via impacts and the accretion of planet embryos and planetesimals; in the Nice model \citep{Gomes2005,Nesvorny2011,Morbidelli2018} the dissipation of the gas disk triggers a dynamical instability, spreading the solids in the system; and in the Grand Tack scenario \citep{Walsh2011,Pierens2014}, Jupiter and Saturn migrate inward during the gas phase until their capture in resonance, inducing an outward migration of the two giants. According to all these models, different noticeable features would be created in the protoplanetary disk as the models require different giant planet configurations, which influence the dust distributions in the disk. The goal of this study is to determine what kinds of features a Solar System embedded in its natal gaseous and dusty disk could produce and how they would be observed by today's instruments. By comparing the resulting images to current observations, we are able to situate the Solar System protoplanetary disk in the wide spectrum of planet-forming disks.

In order to derive these synthetic images, we divided our study into four different steps: we start by simulating the gaseous disk containing the different giant planets using 2D isothermal hydro-simulations with the FARGO2D1D code \citep{Crida2007}. These simulations allow us to derive a detailed radial gas profile that is used as input for a dust evolution model, derived in the TWO-POP-PY code \citep{Birnstiel2012}. This radial dust model takes growth, fragmentation, and drift into account. The resulting dust size distributions are then extended in three dimensions and used in a radiative transfer code, RADMC3D \citep{Dullemond2012}, to derive the synthetic images at different wavelengths; these images are finally convolved with different beam sizes in order to represent more realistic observations.

This paper is structured as follows. First, in Sect. \ref{section_numericalsetups} we outline the different numerical setups of the steps listed above. We present the results in Sect. \ref{section_results}: the results of the dust evolution model are presented in Sect. \ref{section_dustevol}, and then the derived images, convolved with beams, are shown in Sect. \ref{section_images}. The derived images are discussed in Sect. \ref{section_discussion}, and we summarize and conclude our results in Sect. \ref{section_conclusions}.

\section{Numerical setups}
\label{section_numericalsetups}

In this study, we want to simulate how the dust would be distributed in a protoplanetary disk where multiple fully formed giant planets are embedded. We therefore proceed in four steps: i) first, hydrodynamical simulations are run with FARGO2D1D\footnote{\url{http://fargo.in2p3.fr/-FARGO-2D1D-}} in order to determine the gas distribution in the disk, considering different planet configurations, described in Sect. \ref{section_config} and exploring different disk parameters; ii) using the time and azimuthally averaged gas distribution from the hydrodynamical simulations, we investigate how dust behaves in such disks using a dust evolution code called TWO-POP-PY\footnote{\url{http://birnstiel.github.io/two-pop-py/}}, giving us the dust surface density distributions as a function of radius and grain size; iii) we derive the synthetic images from a 3D extension of the dust distributions using the radiative transfer code RADMC3D\footnote{\url{https://www.ita.uni-heidelberg.de/~dullemond/software/radmc-3d/}}, which outputs can then be convolved with different beams in step iv), giving us realistic images of the disks.

In this section we present the different setups for each step, starting by presenting the different planet configurations explored. In Sects. \ref{section_hydro_setup} and \ref{section_dustevol_setup}, we present the numerical setups we used for the hydrodynamical and dust evolution simulations. The radiative transfer code as well as the beam convolutions are presented in Sect. \ref{section_images_setup}.

\subsection{Planet configurations} 
\label{section_config}

As we simulate planetary systems that are still embedded in their gas disks, we consider two possible formation scenarios. Assuming that the four giant planets in the Solar System are already formed (i.e., their mass are fixed to their present value), they are placed in either a Compact or a Spread configuration (see Table \ref{table:1}). In the Compact configuration the planets are located in a tight configuration corresponding to the initial configuration needed by the Nice instability to occur \citep{Gomes2005}; whereas in the Spread configuration, the planets semimajor axis are increased by 30$\%$ compared to their nowadays positions. This aims to take into account migration, assuming that after they formed they migrated inward from further away orbits toward their current configuration \citep{Bitsch2015,Sotiriadis2017,Pirani2019,Oberg2019}.

To investigate planet formation in a global scale, we also chose a third configuration representing an exoplanetary system. This system is studied to help the comparison between the resulting images and the observations. It is composed of three giant planets of 1 Mj located at $\sim 5 \; \rm R_{H,mut}$ \citep{Chambers1996} from each other in the inner region of the disk. This configuration represents a possible intermediate step of N-body simulations aimed to study giant planet formation \citep{Bitsch2020}, but serves also as initial conditions for N-body simulations aimed to explain the eccentricity distribution of giant planets via scattering \citep{Juric2008,Raymond2009a}.

\begin{table}[t]            
\centering                          
\caption{Semimajor axis and masses of the three different configurations considered here (Compact, Spread, and Three Giants).} 
\begin{tabular}{|c c c c c|}        
\hline             
Config & Jupiter & Saturn & Uranus & Neptune \\    
\hline                   
    Compact $\rm r_p$ & 5.45 AU & 8.18 AU & 11.5 AU & 14.2 AU \\ 
    Spread  $\rm r_p$ & 6.76 AU & 12.4 AU & 24.9 AU & 39   AU \\
    Masses            & 1 Mj    & 0.3 Mj  & 0.044 Mj & 0.051 Mj \\
    
\hline\hline
 Config & Giant 1 & Giant 2 & Giant 3 & \\
 \hline
    3 Giants $\rm r_p$ & 5.2 AU & 8.11 AU & 12.69 & \\
    Masses             & 1 Mj   & 1 Mj    & 1 Mj  & \\
\hline
    
\end{tabular}
\label{table:1}
\end{table}

As our goal is to constrain giant planet formation, we also investigate the impact of the planet masses on the disks. By changing the planet masses, we probe different stages in the formation process. We study this effect for two of our configurations, the Spread and the Three Giants, with reduced planet masses corresponding to one-half or one-third of their current mass. In return, we enhance the disk gas masses by 1\% and 2$\%,$ respectively, as the disks harboring the less massive planets represent an earlier evolution step. Only one set of disk parameters is used in this case ($\alpha = 10^{-4}$ and $h = 0.025\times r_{\rm AU}^{2/7}$; see next subsection).

\subsection{Hydro-dynamical setup}
\label{section_hydro_setup}

In order to derive the gas disk profile in presence of four giant planets, we run hydrodynamical simulations with the FARGO2D1D code \citep{Crida2007}. This code is composed of a linear 2D ($\rm r,\phi$) grid, where the planets are located, surrounded by two linear azimuthally symmetric 1D grids. It allows us to simulate the whole viscous evolution of the disk at a reasonable computational cost. The 2D grid spans from $1$ AU to $52$ AU and is prolonged by the inner 1D grid from $0.1$ to $1$ AU and by the outer 1D grid from $52$ to $160$ AU, except when otherwise specified. The interfaces between the 2D and 1D grids are chosen to be far enough from the planets so that we can consider the disk axisymmetric. 


The disk is locally isothermal and subject to an $\alpha$ viscosity as described in \cite{Shakura1973}. We investigate the influence of viscosity by taking three different values for $\alpha = 10^{-4}, \; 10^{-3}, \; 10^{-2}$. For the aspect ratio, we use the minimum mass solar nebula (MMSN) profile, derived by \cite{Weidenschilling1977,Hayashi1981}: $h = 0.033\times r_{\rm AU}^{2/7}$. We also investigate the impact of the aspect ratio by taking a lower value: $h = 0.025 \times r_{\rm AU}^{2/7}$. For simplicity in this paper, the first aspect ratio will be called the "MMSN-like" aspect ratio whereas the second one is referred as the "small" aspect ratio. In each case, the snow line (i.e., the location at which water vapor freezes in the disk, T = 160K) is located at 3.3 AU in the MMSN-like case and at 0.9 AU in the small aspect ratio one. Even if the snow lines are located within the simulation domain, we neglect any physical changes that can occur around this location.

The radial extent of our gaseous initial disk (from 0.1 to 160 AU) is consistent with \cite{Lenz2020} and \cite{Kretke2012}, where they derived the possible properties of the Solar System protoplanetary disk taking into account different available constraints. We also chose a gas surface density profile in agreement with \cite{Lenz2020}: $\rm \Sigma (r) = \Sigma_0\times(r/r_{AU})^{-1}$ , where $\Sigma_0 = 836.1 \rm \; g/cm^2$ at $\rm r = 1 AU$. This value was chosen so that the total mass of the disk is $\rm 0.1 M_{\odot}$. Even if this is considered a heavy initial disk, its large radial extent allows us to neglect self-gravity. Furthermore, as FARGO2D1D features open inner and outer boundaries, the mass of the disk will decrease with viscous evolution.

The resolution is such that the innermost planet (i.e., the planet located at $r_p = 5.2$ AU) is resolved by five grid cells within its Hill radius. This leads, for the 2D grid, to a radial resolution of $\rm N_{r,2D} = 707$ cells with an azimuthal resolution of $\rm N_{\phi,2D}=454$ cells. This corresponds to a radial resolution of $N_{r,1D} = 2218$ when considering the whole disk (i.e., the two 1D grids combined with the 2D grid). As at low viscosity some instabilities can be triggered \citep{Klahr2003,Fu2014}, we enhanced the resolution for $\alpha = 10^{-4}$: $\rm N_{r,2D} = 1414$ and $\rm N_{\phi,2D}=906$ leading to $N_{r,1D} = 4436$. In one particular case, the 2D-1D boundary was too close to the outer planet, creating unrealistic features. Therefore, in the Spread configuration case with low alpha ($\alpha = 10^{-4}$) and small aspect ratio, the 2D-1D outer boundary was moved from $52$ to $78$ AU. The radial resolution was adapted in the 2D part to match the radial resolution of the other simulations at low viscosity: here, $\rm N_{r,2D} = 2135$.

The planets are introduced into the disk with a mass-taper function defined as 
\begin{equation}
    m_{\rm taper} = \sin^2{(t/(4 n_{\rm orb})),}
    \label{masstapereq}
\end{equation}where $n_{\rm orb}$ is the number of orbits used to grow from 0 to $m_p$. Here we chose  $n_{\rm orb} = 10$. The disk is integrated for 12 500 orbits at 5.2 AU ($\sim 1.5 \times 10^5$ years) until it adjusts to the perturbations induced by the giant planets. The 2D density profiles after 12 500 orbits can be seen in Appendix \ref{appendix_2Dgas} for each configuration and disk parameter. The disks are then evolved for another 2 500 orbits at 5.2 AU ($\sim 3.0\times10^4$ years). These gas density profiles are used as an input for the dust evolution setup.

\subsection{Dust evolution setup}
\label{section_dustevol_setup}

To derive the dust distributions in the disks, we used the dust evolution code called TWO-POP-PY \citep{Birnstiel2012,Birnstiel2015}. TWO-POP-PY computes the radial motion of grains as well as their growth from an initial dust and gas radial profile. These initial profiles are derived from the hydrodynamical setup presented above. The same radial resolution is used as in the hydro-simulations ($N_{r,1D} = 2218$ for the highest viscosities and $N_{r,1D} = 4436$ for $\alpha = 10^{-4}$).

The initial gas profile corresponds to the azimuthally averaged gas density and velocity profiles averaged in time over 2 500 orbits at 5.2 AU (average taken between t= 12 500 and 15 000 orbits). This time average will smooth the highly perturbed disk, which is a necessary step as we do not simulate the gas and dust evolution simultaneously in 2D, as in more sophisticated simulations \citep{Drazkowska2019}. We discuss this choice in Sect. \ref{section_discussion_timeevol}. The radial gas density profiles can be found in Appendix \ref{appendix_2Dgas}.

The initial dust profile is derived from the gas profile by assuming a dust-to-gas ratio of 0.01. As we consider that all the planets already fully formed in the disk, we need to take into consideration that some planets already reached the pebble isolation mass and therefore are able to block the pebble flux from the outer regions of the disk  \citep{Morbidelli2012,Lambrechts2014,Ataiee2018,Bitsch2018a}. The pebble isolation mass can be estimated from the aspect ratio $h$ of the disk:
\begin{equation}
    M_{\rm iso} \simeq 20 \: \bigg(\frac{h}{0.05}\bigg)^3 M_{\rm Earth}
.\end{equation}
For our different configurations and each disk scale height profile, Jupiter and Saturn are always above the pebble isolation mass. In the Three-Giants configuration, each planet is above the pebble isolation mass. As some of the giants gaps are able to block the pebble flux from the outer disk, we assume that the dust located between the inner edge of the disk and the outer gap edge of the furthest planet that has reached the pebble isolation mass had time to drift inward during planet formation, making the inner disk depleted in dust. Therefore, our initial dust profile can be written as
\begin{equation}
    \Sigma_{d} =
    \begin{cases}
        0                   & \text{if} \; r\leq r_{p,trunc}+2H_{p,trunc} \\
        0.01\times \Sigma_g & \text{if} \; r>r_{p,trunc}+2H_{p,trunc}
    \end{cases}
    \label{initial_dust}
,\end{equation}
where $r_{p,trunc}$ is the semimajor axis of the planet considered to block the dust flux (Saturn in the Solar System cases, the third giant in the Three-Giants case) and $H_{p,trunc}$ is the disk gas scale height at the location of the planet, used to estimate the position of the outer edge of the planet gap \citep{Paardekooper2006}. The capacity of the giant planets to block the small dust flux is dependent on the dust diffusion and on the gap depth, which in turn depends on the disk viscosity. We show in Sect. \ref{section_dustevol} that if the viscosity is high, the gaps are actually not strong enough and dust diffusion is important, allowing dust from the outer disk to flow in and fill the inner disk \citep{Ovelar2016}. Therefore, the inner disk will remain empty only if the viscosity is low enough to block the dust flux from the outer disk.

The model is evolved for 1 Myr, during which we consider that the gas disk does not evolve significantly (i.e., the gas profile is fixed). The impact of time evolution is discussed in Sect. \ref{section_discussion_timeevol}. During these 1 Myr, the grains grow, fragment and drift. The maximal grain size depends on each limiting mechanism: $a_{\rm max} = \rm min(a_{\rm frag},a_{\rm drift},a_{\rm growth})$. When they are limited by fragmentation, the maximal size the grains can reach is written as \citep{Birnstiel2012}
\begin{equation}
    a_{\rm frag} = \frac{2\Sigma_g u_f^2}{3 \pi \alpha \rho_s c_s^2}
    \label{eq_fraglimit}
,\end{equation}
where $\Sigma_g$ is the gas surface density, $u_f$ is the fragmentation velocity, $\alpha$ is the $\alpha$-viscosity parameter, $\rho_s$ is the internal density of the grains, and $c_s$ is the sound speed. The internal density of the grains is taken to be $\rho_s = 1.675 \rm g/cm^3$, matching the Disk Substructures at High Angular Resolution (DSHARP) survey's composition of grains \citep{Birnstiel2018}. Regarding the fragmentation velocities, velocities of around $\rm \sim 10 m/s$ are required in simulations to explain the formation of large planetesimals \citep{Drazkowska2016,Lenz2019}. However, laboratory experiments of dust collisions under physical conditions of protoplanetary disks  show that these velocities cannot be that high ($\rm \sim 1 m/s$) \citep{Musiolik2019,Schneider2019}. \cite{Pinilla2021} investigate which conditions are necessary for planet formation to happen with a low fragmentation velocity. However, it is clear from Eq. \ref{eq_fraglimit} that a lower fragmentation velocity will yield small grain sizes in the fragmentation limit, if the disk's viscosity is large. In order to reach at least millimeter particles observable with ALMA, we thus use larger fragmentation velocities (1, 3, and 10 m/s) for higher alpha values ($10^{-4}, 10^{-3}$, and $10^{-2}$) in order to have a similar fragmentation limit for all simulations.

The grains are also subject to radial drift: as the gas moves on a slightly sub-Keplerian orbit, the grains feel a head-wind pushing them inward in the disk \citep{Weidenschilling1977,Brauer2008}. The drift barrier corresponds to the limiting size where the grains drift faster than they grow:
\begin{equation}
    a_{\rm drift} = f_d \frac{2 \Sigma_d v_K^2}{\pi \rho_s |\gamma| c_s^2}
    \label{eq_driftlimit}
,\end{equation}
where $f_d = 0.55$ is a correcting factor from \cite{Birnstiel2012}, $v_K$ is the Keplerian velocity and $|\gamma| = \rm | dln(P)/dln(r) |$ is the gas pressure profile of the disk, derived directly from our hydrodynamical simulations.

Finally, the maximum size of the grains is limited by their own growth timescale:
\begin{align}
    &t_g = \frac{\Sigma_g}{e_s\Omega_K\Sigma_d}\\
    &a_{\rm growth} = a_0 \times \rm exp(t/t_g),
    \label{eq_growthlimit}
\end{align}
\noindent where $e_s = 1$ is the sticking probability, $a_0 = 2.5 \times 10^{-6}\rm cm$ is the initial size of all the grains in the disk and $t_g$ is the growth timescale.

Knowing the limiting sizes, the code then computes the motion of the grains by taking into account their relative velocities, their diffusion in the disk and wether they are coupled to the gas or not. This coupling can be monitored with the Stokes number. Assuming spherical grains in an Epstein regime near the disk mid-plane, the Stokes number is defined as\begin{equation}
    \rm St = \frac{\pi a \rho_s}{2 \Sigma_g}
    \label{eq_stnumber}
,\end{equation}
where a is the size of the grain. If $\rm St \ll 1$, then the grains are coupled to the gas and follow its motion in the disk. This has an impact on the dust velocities. TWO-POP-PY divides the different grains into two groups depending on their size (large grain and small grain populations). The velocities of the grains are calculated as an average for each group and are defined as
\begin{align}
    &v_{\rm drift} = \frac{c_s^2 |\gamma|}{2 v_K}\\
    &v_i = \frac{v_{\rm r,gas}}{1 + \rm St_i^2} + \frac{2}{\rm St_i + St_i^{-1}}  v_{\rm drift,}
\end{align}
where $i = \{0,1\}$ for each grain population, $v_{\rm drift}$ is the drift velocity \citep{Birnstiel2012} and $v_{\rm r,gas}$ is the radial velocity of the gas. We present the radial gas velocities from the hydro-simulations used here in Appendix \ref{appendix_2Dgas}. Studies have showed that filtering of dust to the inner parts depend on the Stokes number of the dust and the gas properties \citep{Weber2018,Haugbolle2019}, meaning that the TWO-POP-PY approach gives us a first approximation only of how dust is filtered to the inner parts of the disks.  

After 1 Myr of evolution, we reconstruct the full grain size distribution, determining the surface density of each grain size as a function of orbital distance \citep{Birnstiel2015}. The particle grid used by the reconstruction routine logarithmically ranges from $a_0$ to $6\times \rm a_{max}$, $\rm a_{max}$ being the maximum grain size reached after 1 Myr of evolution, resolved with 300 cells. This dust size-density distribution, presented for each disk in Sect. \ref{section_dustevol}, is the final 1D output used to produce the images.

\subsection{Synthetic image setup}
\label{section_images_setup}

The images are derived using the radiative transfer code RADMC3D \citep{Dullemond2012}. Using the 1D dust size-density distribution from TWO-POP-PY, we extrapolate the 3D distribution of the grains in the disk. We assume a volume density following
\begin{equation}
    \rho_d(r,\phi,z,\text{St}) = \frac{\Sigma_d(r,\text{St})}{\sqrt{2\pi}\;H_d(r,\text{St})}\times \exp{\bigg(-\frac{z^2}{2H_d(r,\text{St})^2}\bigg)}
,\end{equation}
where $z = r\cos{\theta}$ ($\theta$ being the polar angle) and $H_d$ is the dust scale height. $H_d$ is derived from the gas scale height, taking into account vertical settling of grains. The grains having a large Stokes number (Eq. \ref{eq_stnumber}) tend to settle toward the mid-plane, while the grains with small Stokes numbers feel the vertical mixing due to turbulence in the disk and are more coupled to the gas. The dust scale height is therefore given by \citep{Birnstiel2010,Pinilla2021}\begin{equation}
    H_{d} = H_g\times \min{\bigg[1,\bigg(\frac{\alpha}{\rm \min{(St,1/2)}(1+St^2)}\bigg)^{1/2}\bigg]}
    \label{eq_hdust}
.\end{equation}

The hydrodynamical and dust simulations are run with a very high radial resolution, specially at low viscosity. In order to add two dimensions to our disks, we need to reduce the overall resolution due to computational limitations. To do so, we interpolated the dust distributions radially, reducing the resolution by a factor of two (four at low viscosity). The new radial resolution, used to derive the images, is therefore of $N_{r} = 1109$ cells. By applying the same method to the grain size grid, we reduce the number of dust grains by 2. The images are derived considering a disk containing 150 grain size bins. This procedure resulted in no major differences compared to simulations with the full resolution. Thanks to this reduced resolution distributions, the disk can be extended over 320 cells in colatitude ($N_\theta = 320$). Such resolution is needed to correctly derive the temperature of the dust settled to the mid-plane. Finally, we used an azimuthal resolution of $N_{\phi} = 4$ cells, as our disks are considered axisymmetric after the hydro-simulations step.

As mentioned in the previous section, the grains are taken to be spherical grains with the DSHARP composition \citep{Birnstiel2018}. To derive the opacities of such grains, we used OpTool, developed by \cite{Dominik2021}. This library derives the opacity for each grain size using the Mie calculation. OpTool computes the full scattering matrices needed by RADMC3D to include the full treatment of anisotropic polarized scattering.

After computing the temperature profiles of each grain size, RADMC3D derives the images at $\lambda = 1.3 \rm mm$, corresponding to band 6 of ALMA. We assume face-on disks at a distance of 140 pc, which is the typical distance of observed protoplanetary disks, such as in the Taurus, Lupus, and Ophiuchus regions \citep{Gaia2018}. We investigate the influence of the inclination of the disk in Sect. \ref{section_images_differentinclinations} by showing how an inclination $i$ of 30, 45 or 60 degrees influences our results. 
In our configuration, RADMC3D uses $10^7$ photon packages and $5\times10^{6}$ scattering photon packages to derive the raw images. These images are then convolved with a Gaussian beam of FWHM $ = 0.04" \times 0.04"$. In Sect. \ref{section_images_differentbeams}, we investigate different beam sizes that are coherent with different ALMA configurations. At a distance of 140 pc, the $0.04"\times0.04"$ beam size corresponds to a spatial resolution of $\rm 5.6 AU \times 5.6 AU$, which corresponds approximately to the semimajor axis of the inner giant in our different configurations. 


\section{Results}
\label{section_results}

\subsection{Dust size distributions from TWO-POP-PY}
\label{section_dustevol}

\subsubsection{Solar System configurations}

\begin{figure*}[t]
   \centering   
   \includegraphics[scale=0.3]{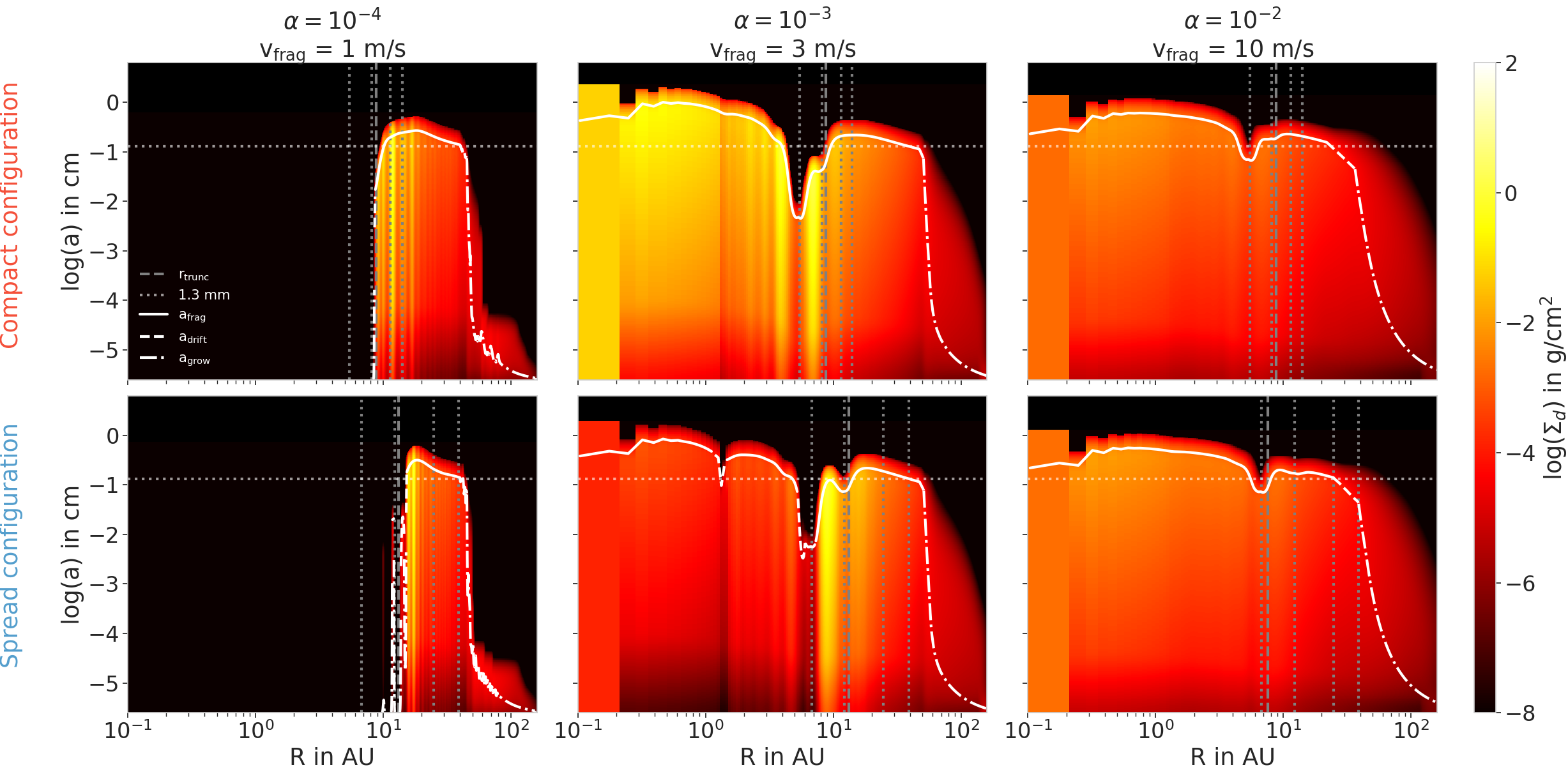}
   \caption{Dust densities distributions for the MMSN-like aspect ratio. Each line represents a Solar System configuration (Compact on the first row and Spread on the second row; cf. Table \ref{table:1}) and each column represents a set of $\alpha$ viscosity and fragmentation speed (increasing from left to right). The vertical dotted lines represent the positions of each planet. The vertical dashed line shows the truncation radius (Eq. \ref{initial_dust}). The white lines represent the maximum size reached by the grains, and each line style represents a limiting mechanism: solid line for the fragmentation limit, dashed line for the drift limit, and dotted dashed line for the growth limit size. At high viscosities, dust from the outer regions flows through Jupiter and Saturn's gap and replenishes the inner disk, which is not the case at low viscosity. At low viscosity, the perturbations induced by the planets in the gas velocity profiles produce dust over-densities (traffic jams), creating substructures not directly related to the positions of the planets.}
   \label{dist_hMMSN}
\end{figure*}

\begin{figure*}[t]
   \centering   
   \includegraphics[scale=0.3]{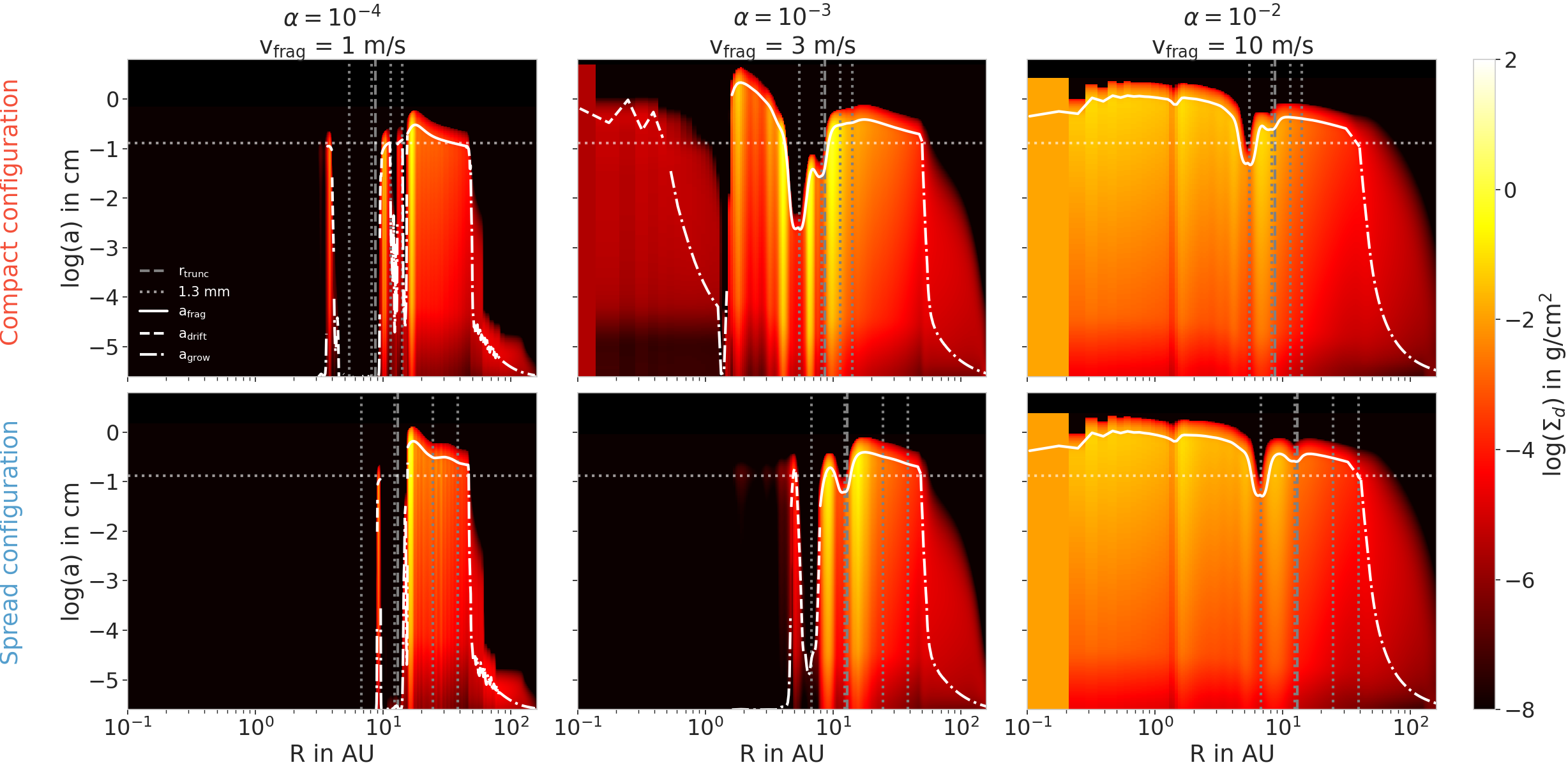}
   \caption{Same as Fig. \ref{dist_hMMSN} but for the smaller aspect ratio. A smaller aspect ratio induces deeper planet-induced gaps in the gas disks, creating stronger features in the dust compared to Fig. \ref{dist_hMMSN}. In the Compact configuration, at $\alpha = 10^{-3}$, a small amount of small dust flows through the gaps of the giants. This produces a low dust-to-gas ratio in the inner region of the disk, resulting in grain sizes limited by growth and drift rather than fragmentation (see Eq.\ref{eq_growthlimit}).}
   \label{dist_h4}
\end{figure*}

In this section we present the radial dust size distributions from TWO-POP-PY after 1 Myr of evolution. In Fig. \ref{dist_hMMSN} we show the distributions in the different Solar System configurations (rows) at different $\alpha$ viscosities and fragmentation speeds (columns) for the MMSN-like aspect ratio. The distributions with a smaller aspect ratio are presented in Fig. \ref{dist_h4}. In both figures,
the white lines represent the maximal size reached by the grains: the solid line shows the part of the disk where the grains are limited by fragmentation, the dashed one corresponds to the drift limit and the dotted dashed line represents the growth limit. We see that these lines are in general above
the dotted horizontal line that marks the 1.3mm size. Vertical dotted lines represent the positions of each planet, and the vertical dashed line shows the location where the dust disk is initially truncated (see Eq. \ref{initial_dust}).

We see that depending on the viscosity, some dust could flow through Jupiter and Saturn's gaps: as expected, at low viscosity $\alpha = 10^{-4}$, the gaps are too deep for the dust from the outer disk to replenish the inner region; however, at high viscosity $\alpha = 10^{-2}$, the dust diffused through the whole disk, leaving no strong substructures. At an intermediate viscosity ($\alpha = 10^{-3}$), we see that depending on the planet configurations, some dust could accumulate around the giant planets locations: in the Compact configuration with an MMSN-like aspect ratio (Fig. \ref{dist_hMMSN}, first row, second panel), dust flows from the outer disk and accumulated between Jupiter and Saturn as well as at the inner edge of Jupiter's gap, creating two over-densities. A similar behavior happened when considering a lower aspect ratio (Fig. \ref{dist_h4}, first row, second panel). However, as a lower aspect ratio implies deeper gaps, Saturn's gap becomes deep enough to accumulate dust at the outer edge of its gap. In this case, the inner disk is more depleted because it is harder for the dust to diffuse through. 

The depletion of dust at low viscosity creates inner cavities. These cavities are observed in several disks \citep{Espaillat2014,VanDerMarel2018} and are described as wide regions in disks where there is no emission observed and therefore possibly no dust present. These cavities can be explained by the presence of planets: either one giant planet is large enough to block the dust flux from the outer disk and the inner disk empties by radial drift; or multiple planets are present and large enough to create a wide common gap. In our simulations, the cavities are created by either mechanism, or a combination of them, depending on the configuration. Either way, the position of the cavity is linked to the position of Jupiter and Saturn, which are both located within 15 AU. In the next section (Sect. \ref{section_images}) we investigate if the resolution of ALMA is sufficient to see the cavities in our cases. 

Comparing the Compact and Spread configurations distributions at $\alpha = 10^{-3}$ for each aspect ratio, we see that there is less dust flowing to the inner disk in the Spread configuration than in the Compact one. This can be explained by the quasi-common gap created by Jupiter and Saturn in the Compact configuration: as Saturn is located further in, the dust has to diffuse through one gap while it has to diffuse through two distinct gaps in the Spread configuration. \cite{Haugbolle2019} found a similar behavior: the presence of a common gap containing Jupiter and Saturn makes filtering less efficient.

Focusing on the Compact configuration with a small aspect ratio for a medium viscosity (Fig. \ref{dist_h4}, first row, second panel), the inner disk is not completely depleted in dust and limited by growth and drift. Reducing the aspect ratio influences the dust behavior in two ways: first, the planet gaps are deeper \citep{Crida2006} and therefore filter dust more efficiently. Moreover, diffusion is reduced as the viscosity is lower for lower aspect ratios. It is therefore harder for the dust to flow through Jupiter and Saturn's gaps. As only some part of the small dust manage to reach the inner part of the disk, the dust-to-gas ratio is very low. As a result, the growing dust particles are limited by growth in the inner region before they drift away, in contrast to the simulations where a lot of dust can reach the inner disk, resulting in grain sizes limited by fragmentation. The capacity of the planet gaps to filter out larger dust can therefore produce an inner cavity with very low surface densities in dust, without being directly linked to the positions of the planets in the disk.

In each panel of both figures, the white lines show that all the disks are roughly fragmentation-limited at radii $r < 50 \rm AU$ and are growth-limited at larger radii. This is due to the fact that the timescales are longer at larger radii (differential rotation). This growth limit sets the size of the disks in the millimeter dust to be $\sim 50 \rm AU$. This size is dependent on the time length of the dust simulation: the millimeter dust disk starts by increasing in size until radial drift reduces the millimeter dust disk significantly. We discuss the impact of time evolution later in Sect. \ref{section_discussion_timeevol}.

At the connection between the growth limit and fragmentation limit ($\sim 50$AU), in each case, we observe a depletion of small dust ($a \lesssim 10^{-4}$ cm). This depletion is due to a narrow region where the dust is actually drift-limited, between the growth and fragmentation limits. We clearly see this regions at high viscosity (dashed line around 30 AU). This drift-limited region makes the largest grains drift inward, before becoming fragmentation-limited. This induces a depletion in small dust as there is no mechanism to replenish this region after the growth of the small dust. This phenomenon was studied by \cite{Birnstiel2015} and creates a gap in the small dust that is not linked to planets at all, but rather to grain growth and drift.

In the Compact configuration at low viscosity and MMSN-like aspect ratio (Fig. \ref{dist_hMMSN}, first row, first panel), we clearly see over-densities that are not linked directly to the gaps caused by planets. These over-densities, not linked to pressure bumps in the gas disk, are due to the highly perturbed gas velocities. As the gas is highly perturbed by the presence of multiple planets (see Appendix \ref{appendix_2Dgas}), the small dust coupled to the gas undergoes "traffic jams": the change of velocity creates over-densities as the dust is slowed down. In these cases, the dust is not trapped and continues to flow after staying in the traffic jam. However, this dust caught in a lower velocity region has time to grow, creating over-densities over a wide range of different sizes. These over-densities are therefore indirectly linked to the presence of the planets and can be observed depending on the resolution of the instruments (see Sects. \ref{section_images} and \ref{section_images_differentbeams}).

\subsubsection{Influence of planet mass and Three-Giants configuration}

\begin{figure*}[t]
   \centering   
   \includegraphics[scale=0.31]{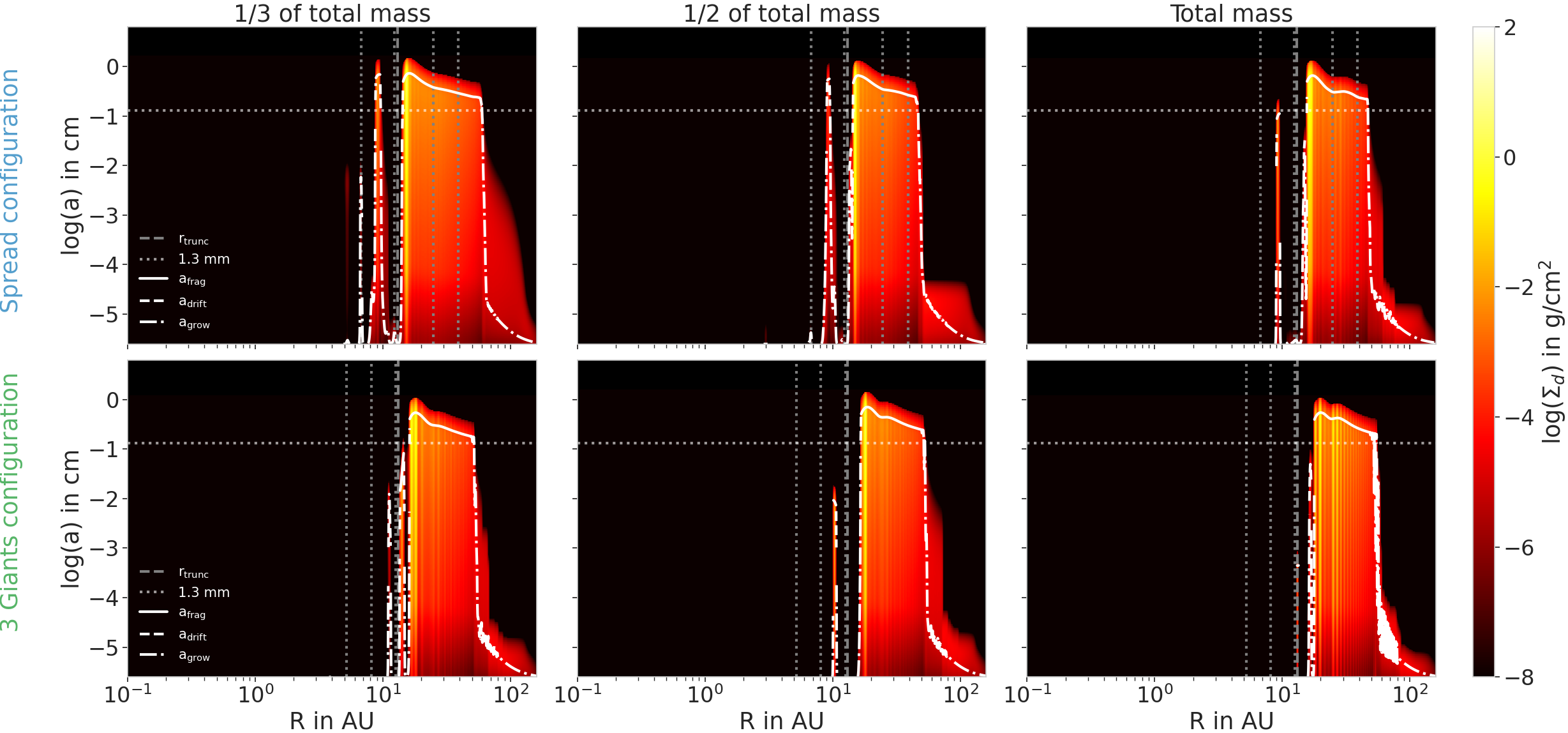}
   \caption{Dust densities distributions with the planet masses reduced by one-half and two-thirds in the Spread configuration (top row) and in the Three-Giants configuration (bottom row). In these simulations, the disk has the smaller aspect ratio and $\alpha = 10^{-4}$ with $v_{\rm frag} = 1 \rm m/s$. The masses of the planets mainly change the gap shapes, allowing more or less dust to flow to the inner regions. In the Three-Giants configuration, we also see that more massive planets create more traffic jams in the disk and therefore present more substructures.}
   \label{dist_diffmass}
\end{figure*}

As presented in Sect. \ref{section_config}, we also investigate the impact of different planetary masses on the dust distributions in the Spread and in the Three-Giants configurations. In the first row of Fig. \ref{dist_diffmass} we show the dust distributions in the Spread configuration case, at low viscosity and small aspect ratio, where the masses are reduced by a factor of two-thirds (left panel) and one-half (middle panel). We can compare them to the total mass case, presented on the right panel. On the second row, we present the distributions of the Three-Giants case, with the different masses as mentioned above. 

Changing the mass of the planets will have two large impacts. The first comes from the initial gas distribution: with increasing planetary mass, the gas is pushed away from the planet more efficiently, depleting the gas disk \citep{Bergez2020}. Our simulations start with lower disk masses for more massive planets to depict the effect of disk evolution during planetary growth. Consequently, as the initial dust content is derived from the gas profile (dust-to-gas ratio of 0.01), the dust disk in the case of full planetary masses is less massive than in the other two cases. As consequence, the maximal grain size reduces for the simulations with increasing planetary mass. Having a more massive dust disk also has an impact on the size of the dust disk in millimeter grains, as it allows faster growth in the outer regions of the disks.

Furthermore, the less massive planets cause shallower gaps. As the gaps are less deep, more dust can flow through the different gaps. For the Spread configuration, the planetary masses results in narrower gap in the dust profile at Saturn's location, while Jupiter is still massive enough to prevent efficient dust diffusion. Regarding the Three-Giants case, reducing the planet masses does not alter the formation of an inner cavity. Even if some dust diffuses through the gap of the third planet in the case of smallest planetary mass, the presence of the other giants is sufficient to keep an inner cavity. Therefore, as expected, bigger planets will create bigger cavities.

On the other hand, planets of different masses will perturb the gas disk differently. More massive planets will induce more perturbations in the gas and create more traffic jams (see the velocity profiles in Appendix \ref{appendix_2Dgas}). This is particularly clear in the Three-Giants case, where we see on the lower panels of Fig. \ref{dist_diffmass} that the outer disk presents different over-densities depending on the planet masses. 

These dust distributions show that each configuration harbors different substructures, mostly at low viscosity. In the next sections we show that some of these features are observable with ALMA.

\subsection{Synthetic images - RADMC3D outputs convolved with Gaussian beams}
\label{section_images}

The dust distributions studied in the previous section present different features, unique to each configuration. In this subsection we present the images and their the radial profiles derived following the setup presented in Sect. \ref{section_images_setup}. First, we focus on the radial profiles at $\lambda =$ 1.3mm for the Solar System configurations at each aspect ratio (Sect. \ref{section_solarsystemprofiles}). The corresponding images can be found in Appendix \ref{appendix_images}. Then we present the images and profiles in the Three-Giants and Spread configurations with the different planetary masses (Sect. \ref{section_diffmassimages}). Different disk inclinations are investigated in Sect. \ref{section_images_differentinclinations} before studying the influence of the beam size on the observable features in Sect. \ref{section_images_differentbeams}.

\subsubsection{Radial profiles in the Solar System configurations at $\lambda = 1.3mm$}
\label{section_solarsystemprofiles}

\begin{figure*}[t]
   \centering   
   \includegraphics[scale=0.29]{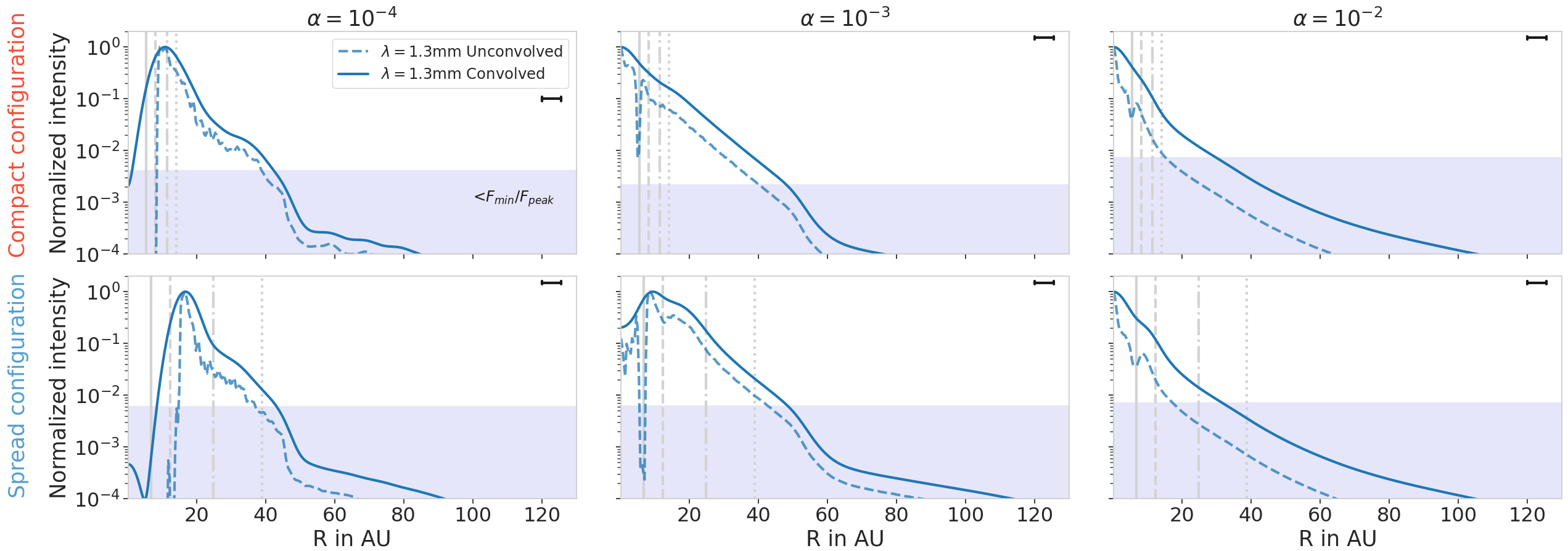}
   \caption{Radial profile of convolved and unconvolved images at $\lambda = 1.3\rm mm$ with the MMSN-like aspect ratio. Each row represents a Solar System configuration, and each column represents an $\alpha$ viscosity. The solid lines show the radial intensity of the images normalized to the peak intensity after convolution with a beam of FWHM = 0.04"$\times$0.04" = 5.6 AU$\times$5.6 AU. The beam is represented with a black horizontal line in the upper-right corner of each panel. The dashed lines represent the normalized intensity of the unconvolved image. Vertical lines show the positions of the planets in each configuration. The light blue area shows the region where the normalized intensity is smaller than $F_{min}/F_{peak}$, where $F_{min} =10 \rm \mu Jy$ is the minimal flux considered to be observable. The value of $F_{min}/F_{peak}$ is different from each panel as $F_{peak}$ is unique to each image, whereas $F_{min}$ is fixed. Comparing the profiles in the case of the convolved and unconvolved images shows us how many features can be missed due to a too low resolution. Regarding the Compact configuration, all substructures are smeared out in the beam.}
   \label{profile_hMMSN}
\end{figure*}

\begin{figure*}[t]
   \centering   
   \includegraphics[scale=0.29]{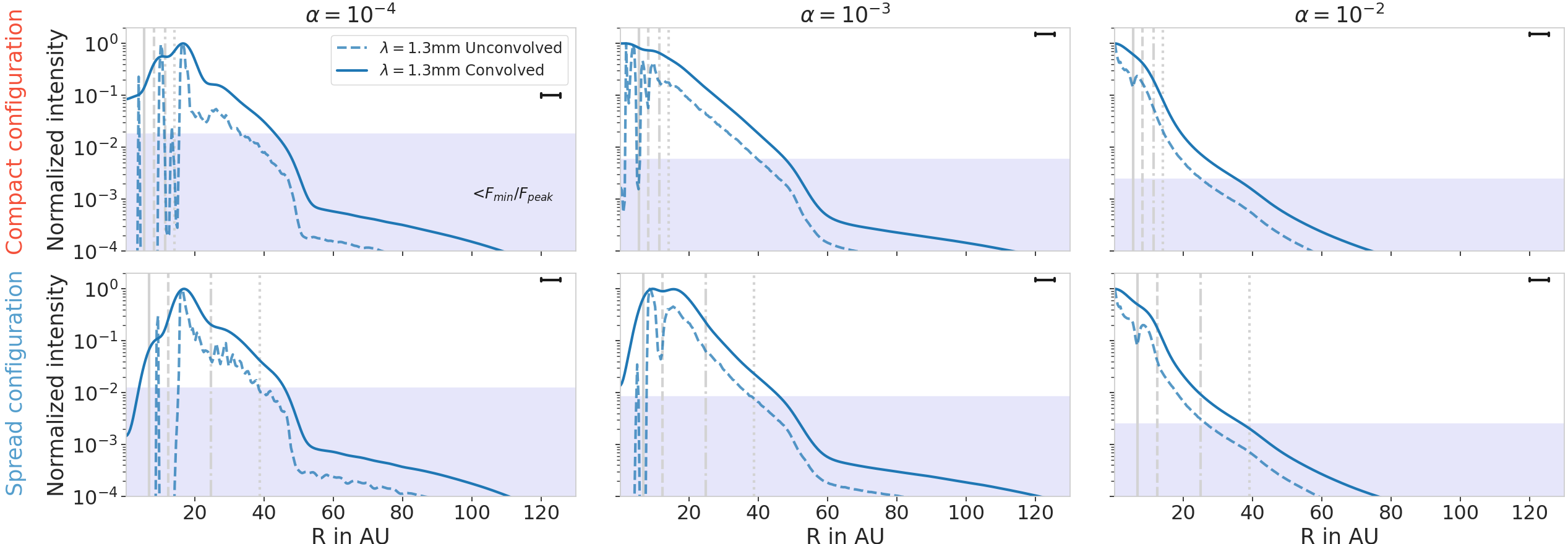}
   \caption{Same as Fig. \ref{profile_hMMSN} but for a smaller aspect ratio. As the dust distributions show more intense substructures, some features become observable but the majority are still smeared out in the beam.}
   \label{profile_hsmall}
\end{figure*}

In order to determine which features are observable, we show in Figs. \ref{profile_hMMSN} and \ref{profile_hsmall} the radial profiles of the normalized intensities (intensity of the image normalized to the peak intensity along one radius of the disk) for images with unconvolved (dashed) beams and for images with a $0.04"\times0.04"$ Gaussian beam convolution (solid). The corresponding images can be found in Appendix \ref{appendix_images}. In the images, we assumed that the minimum flux that can be received due to noise is $F_{\rm min} = 10 \rm \mu Jy/beam$ \citep{Andrews2018}. This minimum flux is represented in the radial profiles by the blue regions: in each profile, this minimum flux is normalized to the peak intensity and emission present in this region can be assumed to be lost in the noise of the images. The value of $F_{min}/F_{peak}$ is different for each panel as $F_{peak}$ is unique to each image, whereas $F_{min}$ is fixed.

As in the previous figures, each row represents a planet configuration and each column represents an $\alpha$ viscosity. It should be noted that for a better comparison with the images, we present the profiles on a linear radial scale, whereas in the previous section the dust distributions are presented along a logarithmic radial scale.

Regarding the high viscosity cases ($\alpha = 10^{-2}$), as expected from the dust distributions, the disks show almost no noticeable feature. In the unconvolved images, we can distinguish the gap created by Jupiter. However, the gap is too close to the star and too small to avoid being smeared out by the beam. Therefore, if the viscosity is too high, then a Solar-System-like planetary structure would be completely invisible in the dust disk. This is consistent with the work of \cite{deJuanOvelar2016}, where they show that a high viscosity disk does not present strong substructures. It is also consistent with \cite{Zhang2018} where they show in their Section 5.1 that a Solar System protoplanetary disk featuring our giant planets in nowadays configuration would not present strong substructures if the viscosity is too high.

\begin{figure*}[t]
   \centering   
   \includegraphics[scale=0.29]{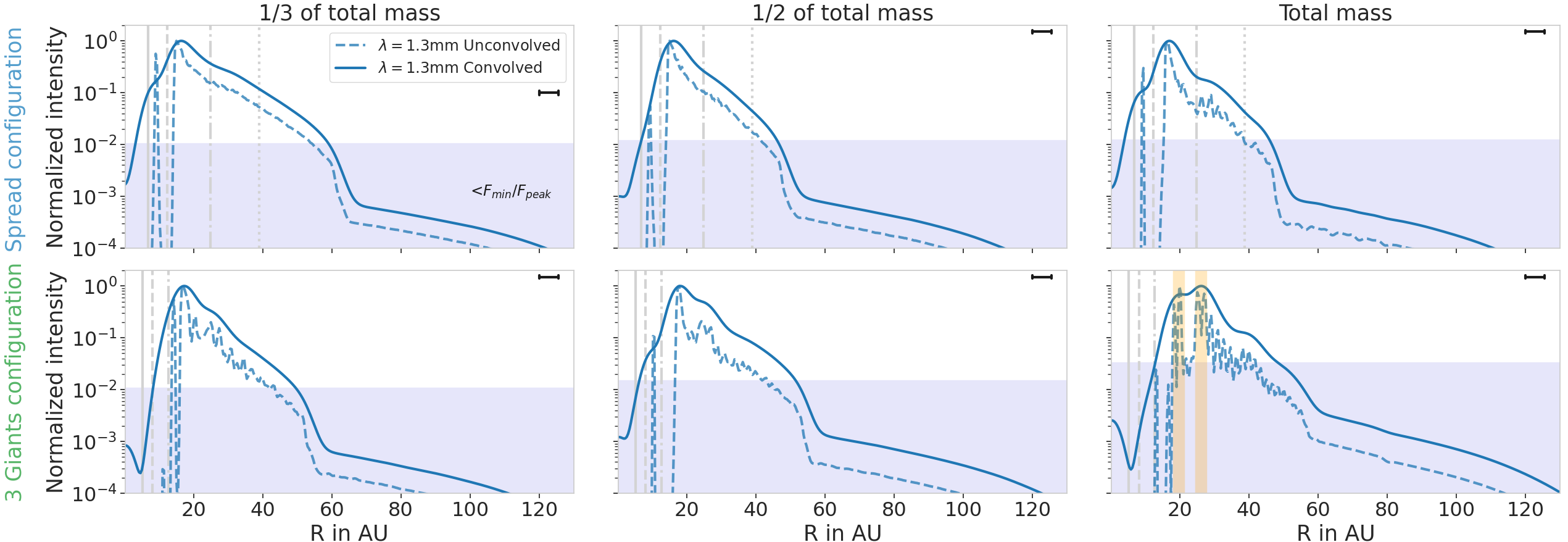}
   \caption{Radial profile of convolved and unconvolved images at $\lambda = 1.3\rm mm$ with the planet masses reduced by one-half and two-thirds in the Spread configuration (top row) and in the Three-Giants configuration (bottom row). In these simulations, the disk has the smaller aspect ratio and $\alpha = 10^{-4}$ with $v_{\rm frag} = 1 \rm m/s$. The intensity profiles are presented as in Fig. \ref{profile_hMMSN}. The convolved profiles are convolved with a beam of 0.04"$\times$0.04" = 5.6 AU$\times$5.6 AU, represented with a black horizontal line in the upper-right corner of each panel. One can notice the similarities between each convolved profile, making it difficult to disentangle between each evolutionary stage and configuration. In the Three-Giants configuration, we highlight in orange the substructures outside of the giants' region, originating in the traffic jams discussed in Sect. \ref{section_dustevol}.}
   \label{profile_diffmasses}
\end{figure*}

\begin{figure*}[t]
   \centering   
   \includegraphics[scale=0.3]{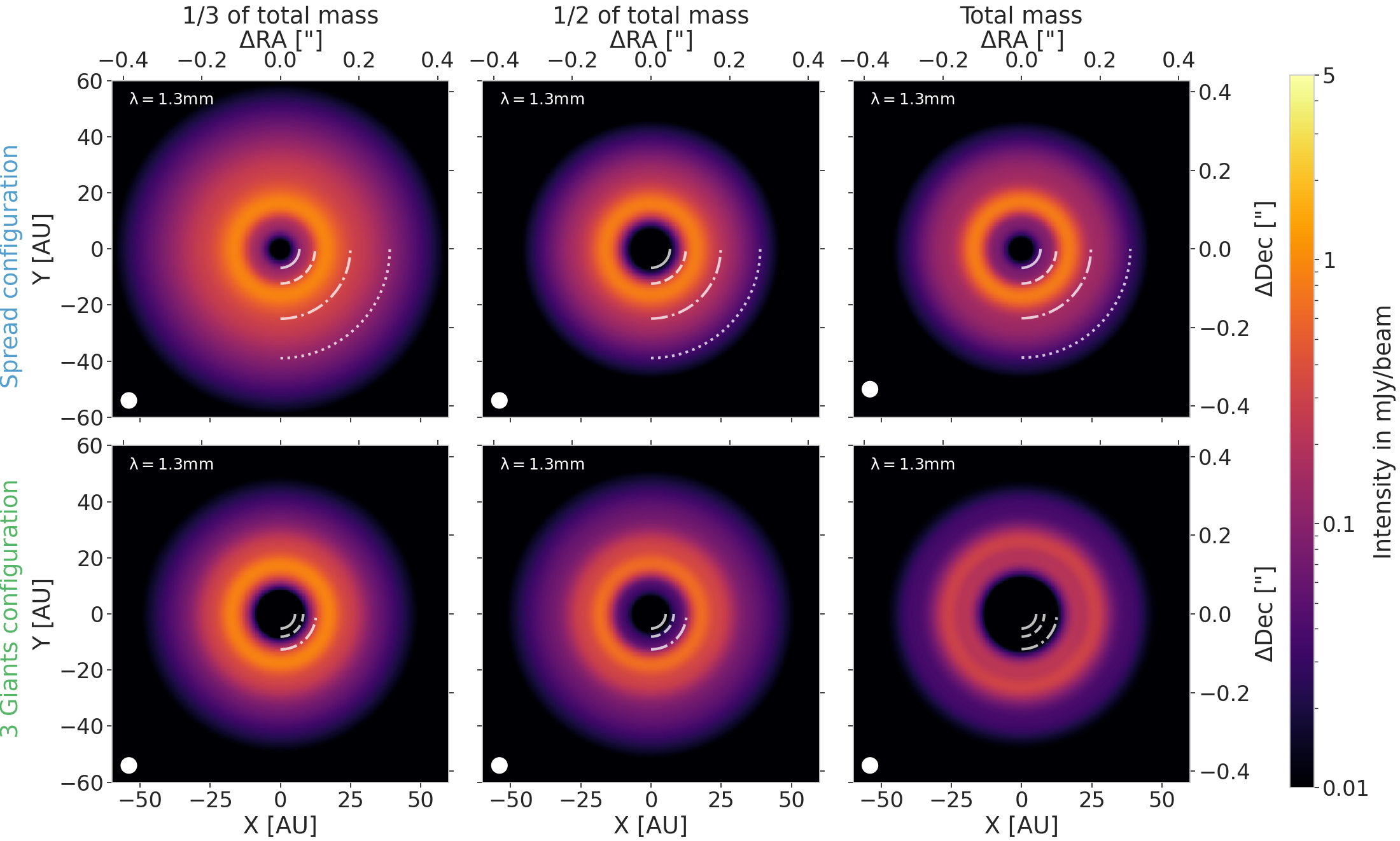}
   \caption{Images in total intensity at $\lambda = 1.3\rm mm$ corresponding to the radial profiles presented in Fig. \ref{profile_diffmasses}. The positions of the different planets are represented by the different white arcs. The beam size ($0.04"\times0.04"$) is represented in the lower-left corner of each image by the white ellipse. In the Spread configuration, we see that the size of the disks depends on the masses of the planets. In the Three-Giants configuration, one can notice the substructures outside of the giants' region corresponding to the orange regions of the profiles in Fig. \ref{profile_diffmasses} and originating from the traffic jams observed in Sect. \ref{section_dustevol}; these traffic jams were created by the perturbed velocity profile of the disk (see Fig. \ref{vrad1D_diffmass}).}
   \label{image_1300m_diffmasses}
\end{figure*}

The images at $\alpha = 10^{-3}$ show a similar pattern: the features are mostly either too small or too close to the star to be distinguishable with this resolution. Although, in the Spread configuration case, the inner disk starts to be depleted in dust (see Figs. \ref{dist_hMMSN} and \ref{dist_h4}), resulting in a decrease in the normalized intensity in the inner regions of the images. These small inner cavities are located at a radius within Jupiter's orbit: the giant planets are therefore outside the cavity in this case. In this same configuration, Saturn's gap start to be large enough to induce a small dip in the intensity profile, slightly noticeable in the convolved images.

At low viscosity ($\alpha = 10^{-4}$), the dust distributions showed strong inner cavities and several substructures. The convolution with a beam of this size kept the inner cavities in all configurations, even if they are deeper in the Spread configuration than in the Compact one. In all the configurations, the intensity profiles decrease rapidly around 50 AU, which is caused by the growth limit of the grains discussed in the previous section. As the millimeter size dust is the dust that contributes the most to the 1.3 mm emission, the growth limit sets the size of the disk in the images (as can be seen in Appendix \ref{appendix_images}). In the Spread configuration, depending on the sensitivity of the instrument, Neptune can be located close to the edge of the disk but only the growth limit sets the location of the drop of intensity. 

As discussed in the previous section, the multiple planets can create some substructures in the disk not directly linked to the positions of the planets. Even if these over-densities can be seen in the unconvolved images, the beam smeared out the majority of them. However, in the Compact case, low viscosity, low aspect ratio (first top panel of Fig. \ref{profile_hsmall}), we see several dips in the intensity profile. The first one is linked to the inner cavity (r < 10 AU). The second one is located at Uranus and Neptune orbits and originates from the small gaps that the two icy giants create in the gas and dust disk. However, the two gaps are indistinguishable here due to the beam size, reducing the emission of the dust located between the two planets. Similarly, as some dust piled up at the outer edge of Neptune's gap and due to the shape of the fragmentation limit in this case, a small part of the disk is shadowed outside of Neptune's orbit, creating a third dip in the intensity profile.

Focusing on the Spread configuration at low viscosity ($\alpha = 10^{-4}$) at each aspect ratio, Figs. \ref{profile_hMMSN} and \ref{profile_hsmall} show a bump in the intensity between the orbits of Saturn and Uranus. This bump originates from the pileup of dust that is blocked at Saturn's outer gap edge. This bump creates a bright ring separating the inner giants and the icy giants, located around 15 AU. This configuration and viscosity is the only setup that presents a bright clear ring at this resolution. We discuss this peculiarity in Sect. \ref{section_comparison_ALMA}, where we compare our disks to known observed disks with similar resolutions.

In summary, the Solar System configurations do not present a lot of substructures in general at this resolution. At low viscosity, the Compact and Spread configurations are presenting very different features, with the Spread configuration showing a clear bright ring between Saturn and Uranus, whereas the Compact configuration presents features that are not directly linked to the positions of the planets. Therefore, the detectability of substructures is highly dependent on the disk viscosity and planet configuration.

\subsubsection{Influence of the different masses on the 1.3 mm images}
\label{section_diffmassimages}

As discussed previously, the planet masses have a lot of impact on the dust distributions, creating different features in the disks. We show in Fig. \ref{profile_diffmasses} the radial normalized intensity profiles in the simulations with different planetary masses, for the Spread and the Three-Giants configurations. Due to the different masses of the dust disk, they have different sizes: in the Spread configuration (top row), we see that the drop in intensity due to dust growth is located at different radii. This effect is even more noticeable in Fig. \ref{image_1300m_diffmasses}, where we show the disks as they would be observed. This effect is less present in the Three-Giants case because the planets are located in the very inner region, therefore not having a strong impact on the gas distribution in the outer regions of the disk (see Appendix \ref{appendix_2Dgas}).

As noticed in the previous section, having planets of different mass influence the amount of dust that can flow through the gaps of the giants. This has a very small impact here as the beam is too large to resolve the small amount of dust present between Saturn's gap (or the third giant's gap in the Three-Giants case) and the inner cavity. The only case where enough dust managed to flow to the inner regions is in the reduced by two-thirds case in the Spread configuration, as Saturn is only slightly above pebble isolation mass. Here, the amount of dust is large enough to slightly enhance the intensity between Jupiter and Saturn's orbit. This slight enhancement is particularly visible in the image in Fig. \ref{image_1300m_diffmasses} (first top panel). On the other hand, as more massive planets block dust most efficiently, it accumulates more at the outer edge of Saturn's orbit. This results in a brighter and clearer ring separating the inner giants from the icy giants.

Regarding the features created by the different traffic jams, most of them are not strong enough to be noticed in the Spread configuration. However, in the Three-Giants case, the traffic jams create rings of different intensities. Expectedly, strongest over-densities create brighter rings. As the strength of the traffic jams are dependent on the planet masses, the most massive case present the strongest features. In the end, these two consecutive rings are due to the gas radial velocity profile and are not directly linked to the presence of planets close to the bright rings.

As already mentioned, changing the planet mass allows us to probe different stages of planet formation (i.e., different times in the formation process). One can notice that the differences between the Spread configuration with half its mass is quite similar to the reduced planet masses in the Three-Giants configuration (Figs. \ref{image_1300m_diffmasses} and \ref{profile_diffmasses}, middle panel of first row compared to the first and second panels of the second row). Therefore, in order to really disentangle planet formation processes, better resolution is needed. We discuss this in Sect. \ref{section_images_differentbeams}. 

\subsubsection{Influence of the disk inclinations}
\label{section_images_differentinclinations}

\begin{figure}[t]
   \centering   
   \includegraphics[scale=0.265]{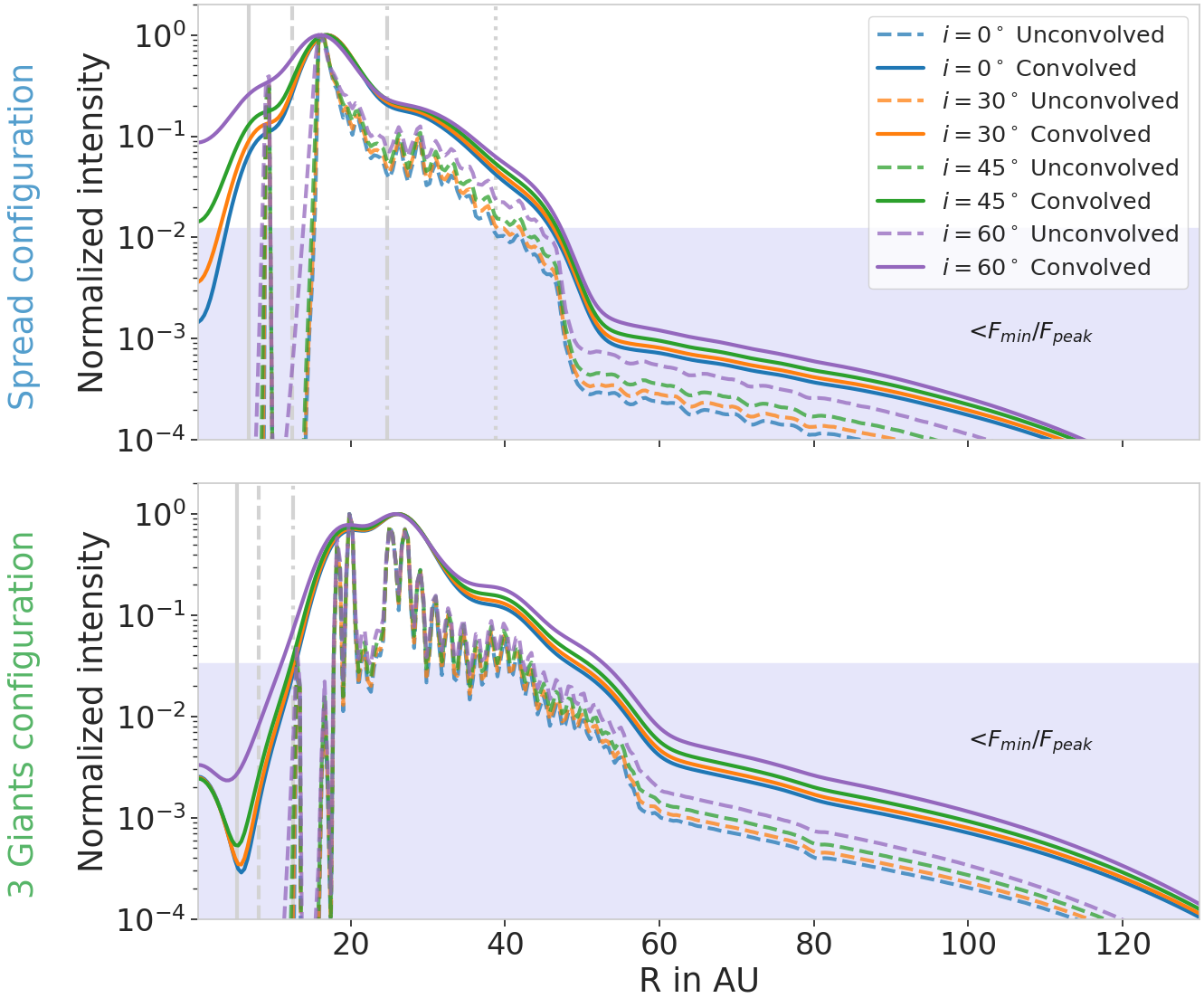}
   \caption{Radial profile of convolved and unconvolved images at $\lambda = 1.3\rm mm$ with different inclinations in the Spread configuration (top row) and in the Three-Giants configuration (bottom row). Here, the disks have the smaller aspect ratio and $\alpha = 10^{-4}$ with $v_{\rm frag} = 1 \rm m/s$. The convolved images present a beam of $0.04"\times0.04"$. The profiles are derived from images that have inclinations ranging from $i = 0^\circ$ to $i = 60^\circ$. The light blue area shows the region where the normalized intensity is smaller than $F_{min}/F_{peak}$ (see Fig. \ref{profile_hMMSN}). The only highly impacted region is the inner cavity: more inclined disks hide the inner regions more efficiently, influencing the depth of the observed cavities.}
   \label{profile_inclinations}
\end{figure}

Many of the observed disks are actually inclined compared to our line of sight \citep{ALMA2015,Andrews2018}. In this section we explore how the inclination of these disks can have an impact on the visible substructures discussed in the previous sections. In order to do so, as in the previous section where we explored the impact of the beam size, we derive the images in the Spread configuration at low aspect ratio and viscosity and in the Three-Giants configuration and infer three different inclinations to the disks: $i = 30, 45, 60^\circ$. The images at different inclinations can be found in Appendix \ref{appendix_images_inclined}.

In Fig. \ref{profile_inclinations} we present the radial profiles of the normalized intensity with different inclinations. The radial profiles are taken to be a section of the image along the semimajor axis of the inclined image. No deprojection procedure was applied, as the disks are axisymmetric by construction. By taking the profile along the semimajor axis, we look at the section of the disk that is situated at the same distance from the observer, independently of the inclination. As in the previous profiles, we show the unconvolved profiles with dashed lines and convolved profiles with solid lines. 

The profiles are very similar, presenting the same features in each case. The main difference resides in the inner cavities: a more inclined disk will hide the depth of the inner cavity as the dust present closer to the observer (lower part of our images) will hide the cavity. However, the cavities are wide enough to be visible: if the cavity is too small, the dust from the closer part of the disk will completely hide the cavity to the observer. On the other hand, larger cavities will be less impacted by the disk inclination. This can be seen with our configurations: in the Spread configuration, the inner cavity is smaller than in the Three-Giants one (see Figs. \ref{dist_diffmass} and \ref{image_1300m_diffmasses}) and the inclination of the disk has a stronger effect on the inner cavity in the Spread configuration.

\begin{figure*}[t]
   \centering   
   \includegraphics[scale=0.285]{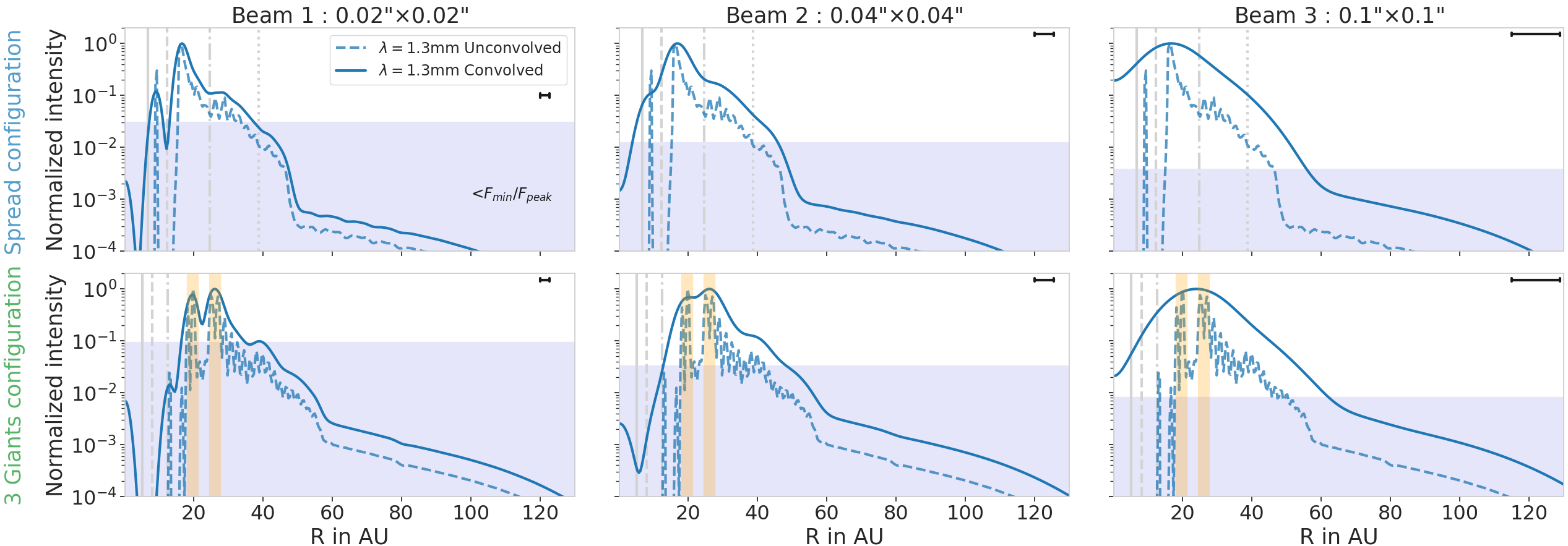}
   \caption{Radial profile of convolved and unconvolved images at $\lambda = 1.3\rm mm$ with different beam sizes in the Spread configuration (top row) and in the Three-Giants configuration (bottom row). In these simulations, the disk has the smaller aspect ratio and $\alpha = 10^{-4}$ with $v_{\rm frag} = 1 \rm m/s$. The different beams investigated range from $0.02"\times0.02"$ = 2.8 AU$\times$2.8 AU (left) to $0.1"\times0.1"$ = 14 AU $\times$ 14 AU (right). We present the fiducial resolution ($0.04"\times0.04"$ = 5.6 AU $\times$ 5.6 AU) in the middle panel for comparison. Each beam is represented by a horizontal line in the upper-right corner of each panel. Orange regions represent the substructures produced by traffic jams, clearly visible in the Three-Giants configuration. The different profiles show that the highest resolution is really needed to start to correctly represent the features of the dust disk.}
   \label{profile_beams}
\end{figure*}

\begin{figure*}[h!]
   \centering   
   \includegraphics[scale=0.32]{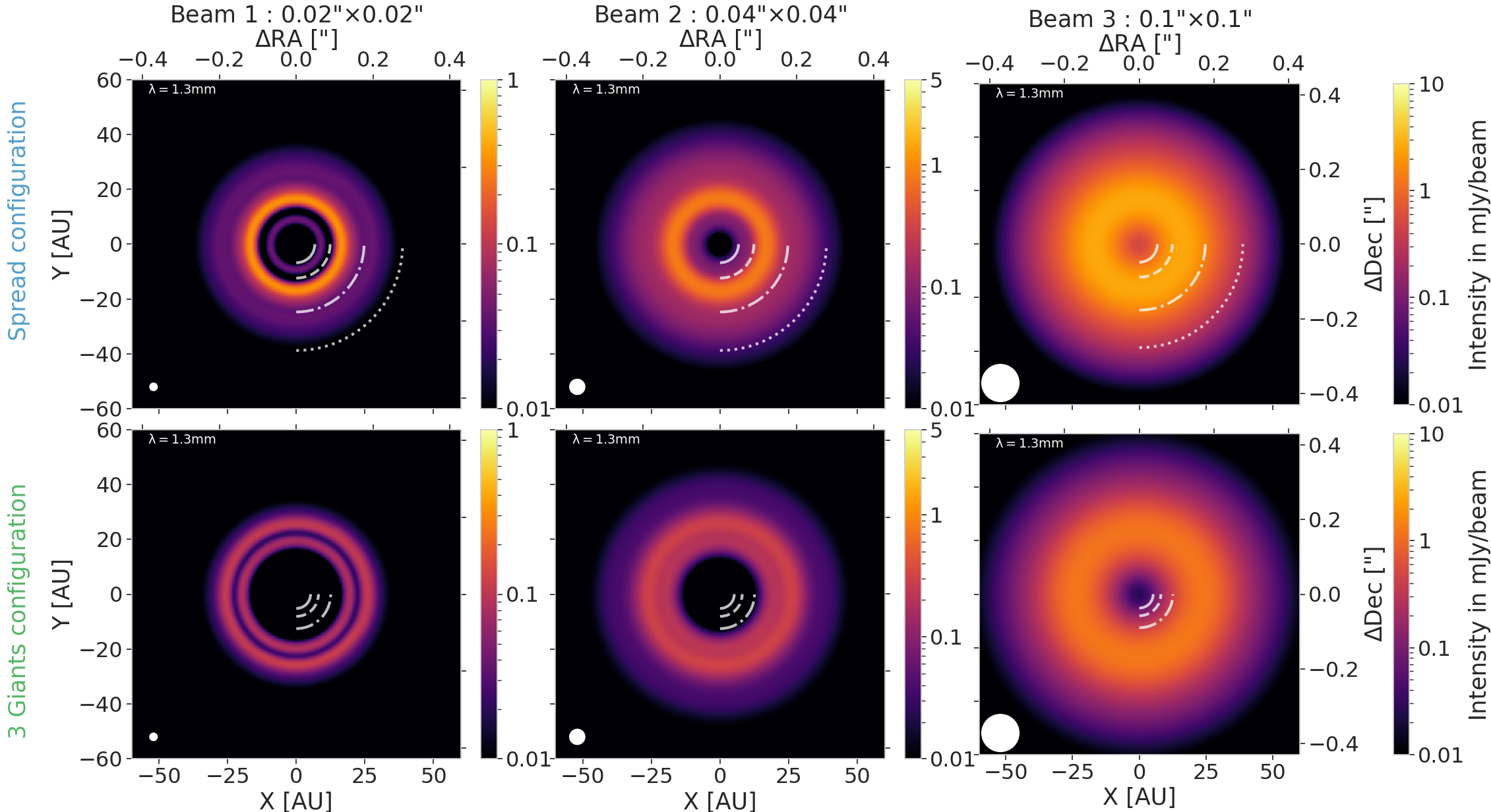}
   \caption{Images in total intensity at $\lambda = 1.3\rm mm$ corresponding to the radial profiles presented in Fig. \ref{profile_diffmasses}. The positions of the different planets are represented by the different white arcs. The different beams investigated range from $0.02"\times0.02"$ (left) to $0.1"\times0.1"$ (right) and are represented in the lower-left corner of each image by the white ellipses. We show the fiducial resolution ($0.04"\times0.04"$) in the middle panel for comparison. The colorbars are adjusted for each resolution as they have different sensitivities. These images show the importance of resolution: the substructures start to be well represented at very high resolution.}
   \label{image_beams}
\end{figure*}

As disk millimeter emissions are geometrically flat \citep{Birnstiel2010,Pinilla2021}, we show here that the inclination of the disk does not have a strong effect on the observed profiles. This means that the inclination of the disk does not hide or create features that could originate from giant planets, important to derive constraints for planet formation.

\subsubsection{Influence of the beam size}
\label{section_images_differentbeams}

ALMA can reach different resolutions depending on the observed wavelength and configuration. As we derived the images at $\lambda = \rm 1.3mm$, we are interested in the Band 6 observations. With the different configurations available, the most common resolutions reached are therefore equivalent to beams of three different size: the most resolved one has a beam of $0.02"\times0.02"$ \citep[as in][]{Benisty2021}, the most common one has a high resolution with a beam of $0.04"\times0.04"$ \citep[as in][]{Andrews2018} (used in the previous sections) and the last configuration gives a beam of $0.1"\times0.1"$ \citep[as in][]{Long2018,Kurtovic2021}. In Fig. \ref{profile_beams} we present the radial profiles of the normalized intensity in the Spread and Three-Giants configurations, with a small aspect ratio and low viscosity. The corresponding images are shown in Fig. \ref{image_beams}. As we present the images in intensity per beam, the sensitivity are different for each resolution: the colorbars range from $F_{\rm min}$ to different maxima depending on the resolution.

In the Spread configuration, we noticed in the previous section that the dust situated between the inner giants and the icy giants creates a bright ring when the resolution is $0.04"\times0.04"$. When the resolution is lower (right panels), the emission is spread over Jupiter and Saturn's orbit, completely hiding Saturn's gap. However, the inner truncated disk is still noticeable, with a small decrease in the intensity. On the other hand, when we compare the high resolution case (at 0.04") with the highest resolution (at 0.02"), we see that the ring is clearly located at the outer edge of Saturn's gap. The resolution is even high enough to start distinguishing Uranus's gap and the small over-density of dust located between Jupiter and Saturn. On the other hand, with a minimum flux situated at $10 \mu \rm Jy/beam$, Neptune is completely missed and lost in the noise.

Regarding the Three-Giants configuration, we see a similar behavior: with a beam of $0.1"\times0.1"$ (right panels), the substructures created by the perturbations by the giants are completely washed out and the only feature remaining is the inner cavity. In both configurations, one can notice that the cavity is shifted compared to the planets orbits: the giants are located in the decrease in intensity, not at the minimum. As we noticed in the previous section, a resolution of $0.04"\times0.04"$ is sufficient to start to distinguish the over-densities of dust located at the outer edge of the further giant gap. The highest resolution is needed to really resolve the main over-densities that are due to the perturbations of the gas velocity by the multiple giants. With this very high resolution, two rings are observed, corresponding to the two brightest set of over-densities seen in the dust distributions (bottom left panel of Fig. \ref{dist_diffmass}).

Changing the resolution has an impact on the observable features, but has also an impact on the observed size of the disk. As the beam spread the intensity in the disk, this enhances the value of $F_{\rm peak}$, reducing the value of $F_{\rm min}/F_{\rm peak}$, as we assume a fixed $F_{\rm min}$ value. Therefore, as the intensity decreases with the radius, the observed size of the disk will depend on the resolution used, presenting a larger disk at lower resolutions. As the disk size can change depending on the image resolution, alternative methods such as uv-modeling \citep{Hendler2020} should be considered to retrieve such a quantity.

\section{Discussion}
\label{section_discussion}

\subsection{Comparisons to known observed disks}
\label{section_comparison_ALMA}

The DSHARP survey \citep{Andrews2018} studied several bright massive disks around stars located in the vicinity of the Sun with a beam size of $\sim 0.035"$. These disks present several features, such as gaps, rings, spirals and asymmetries. The most axisymmetric disks show several configurations of gaps and rings \citep{Huang2018}. Some of the disks present bright rings located at large radii (r > 50 AU), such as AS209 and HD163296. Our synthetic images never showed features at such large radii, even in the Spread configuration where Neptune is located at 39 AU. This can be explained by the sizes of the disks: our gas disk is small (160 AU in radius) whereas HD163296 is thought to be wider than 500 AU in radius \citep{Isella2007,MuroArena2018} with a potential outer planet located at 260 AU \citep{Pinte2018}. 

One of the explanations for the existence of structures at large radii is the presence of planets carving gaps and creating rings. \cite{Lodato2019} show that with giant planets migrating fast enough, it is possible to produce gaps and rings at large radii and still reproduce the distribution of eccentric giant planets observed in radial velocity. However, such migration speeds require a too high viscosity \citep{Ndugu2019} compared to the viscosity needed to allow the formation of planets that could explain the observed substructures. On the other hand, it is also very unclear how planets can form that far in disks with the core accretion model \citep{Morbidelli2020}. Moreover, our images with the Three-Giants configuration do not present features in the outer disk because the planets are located in the inner disk. It is possible that this configuration will lead afterward to some scattering events that will produce systems with giant eccentric planets \citep{Bitsch2020}. We show here that planets forming in the inner disk do not result in features (rings or gaps) in the outer disk region as observed in the DSHARP survey. It is clear these two different giant planet formation channels result in different observable disk structures.

Another important point is linked to the substructures induced by the planets but not directly linked to their orbit and gap. For example, \cite{Zhang2018} analyze the rings and gaps structures present in the DSHARP disks and derive which possible planet mass could produce such substructures. Even if they take into account the fact that some planets can create multiple gaps at low viscosity \citep{Dong2018,Bae2018}, as in AS 209, we found that the gas radial velocity structure can also create rings, blurring even more the link between the number of planets and the number of gaps present in the disk. We discuss in Sect. \ref{section_diskuss_trafficjams} how this problem could be addressed.

However, the disk surveys are biased toward the brightest disks. The differences in the images between the massive DSHARP disks and our study confirm that planet formation happening in the inner regions of the disk results in different features in observed disks. However, these surveys contain some bright disks that are similar in size and show comparable features as the disks studied here. In the Ophiuchus DIsc Survey Employing ALMA (ODISEA) \citep{Cieza2019,Williams2019}, DoAr44 presents a bright inner ring and a dimmer one exterior to it, resembling the image of the Spread configuration disk at low viscosity and low aspect ratio with a resolution of 0.02" (see Fig.\ref{image_beams}, first top panel). Similarly, \cite{Facchini2020} observed two disks, LkCa15 and J1610 showing features similar to our Three-Giants configuration disk with a resolution of 0.04" (see Fig. \ref{image_beams}, middle panel of second row). Observations of the V4046 Sgr circumbinary disk by \cite{Martinez-Brunner2021} present features that are very similar to the Spread Solar System observed with a similar resolution (in Fig.\ref{image_beams}, top left panel), around a binary, unlike our configuration. This images prove that ALMA is capable of reaching such high resolution. Observations of such disks can therefore give us some insights on how planet formation can occur in the inner regions of the disks, compared to the DSHARP observations giving insights on how it occurs in the outer regions of the disks. Moreover, the constrains derived from the local study of the Solar System could help understand planet formation in disks such as DoAr44 that present similar features as our Solar System disks.


\subsection{Comparing to exoplanet populations}
\label{section_comparison_exoplanets}

In Sect. \ref{section_images_differentbeams} we show that with the highest resolution, it is possible to observe features originating from the ice giants if they are in the very outer regions of the disk and far away from the inner gas giants. Microlensing surveys, such as that presented in \cite{Suzuki2016}, claim that the most common type of planets observed are ice giants located at a few AU from their star. Our study shows that it is possible to observe, with the highest resolution that ALMA can reach, features caused by such planets if they are at a few tens of AU. These observations could therefore help to constrain the formation pathways of the ice giants found in microlensing surveys, under the assumption that these ice giants do not turn into gas giants. We should note here that microlensing surveys mostly observe dwarf stars, which should have less massive disks in the first place, making observations unfortunately very difficult. 

Constraining planet formation during the disk phase is important to improve our understanding of different formation scenarios. Indeed, the gas disk phase contains information about the initial conditions of planet formation and the initial structure that could lead to dynamical instabilities after the disk phase. The final structure of the planetary system depends highly on the processes happening during the gas disk phase. 

Currently, the formation of giant planets is still unclear. In one hand, observations of large disks, such as in the DSHARP survey, motivate the idea that giant planets must form in the outer part of the disks and then migrate inward, explaining the presence of bright outer rings and the planet distributions observed by different surveys \citep{Lodato2019}. However, this scenario requires a rather high viscosity in order to have an efficient migration of the giant planets. \cite{Ndugu2019} shows that if the viscosity of the disk is lower, as disk observations seem to suggest \citep{Dullemond2018,Flaherty2018}, then these giant planets do not have time to migrate to semimajor axis corresponding to distances within the reach of radial velocity surveys \citep[e.g.,][]{Fulton2021}.

Another possible giant planet formation scenario is to have giant planets forming in the inner regions of the disks, where the orbital timescales favor planet formation and where a slower migration of the planets can still explain the observed giant distributions \citep{Bitsch2020}. Our study shows that giant planets forming in the inner part of the disks do not produce bright features as observed in the DSHARP survey.

Higher resolution observations of disks can therefore help us distinguish between the formation of planets in the outer disk or in the inner regions of protoplanetary disks. This can give constraints on the initial conditions needed for planet formation to occur and improve the link between the different observed planet populations and the theoretical models studying different planet formation scenarios. 

\subsection{Features created by traffic jams}
\label{section_diskuss_trafficjams}

In Sect. \ref{section_dustevol} we present the dust distributions in each configuration. Some of the distributions show multiple narrow dust over-densities, specially at low viscosity. The configuration showing the clearest over-densities is the Three-Giants configuration (Fig. \ref{dist_diffmass}). These dust rings are created by traffic jams, as shown in Appendix \ref{appendix_gasvrad}. As these traffic jams originate from the highly perturbed gas radial velocities and not from a pressure bump present in the gas, the dust is not trapped and will flow to the inner parts of the disk. 

The presence of these traffic jams has several impacts on our understanding of planet formation. First, as they create features observable by ALMA, it blurs further the link between the number of planets present in disks and the number of gaps and rings created. Considering that a single planet can create multiple gaps in low viscosity disks \citep{Dong2018,Bae2018}, having features created by velocity perturbations on top of the one created by pressure perturbations complicates our estimations of planet masses needed to create observed features.

However, in order to trigger the formation of these traffic jams, the disk needs to be highly perturbed in velocity. In our simulations, it requires the presence of multiple giant planets. In systems where only one or two giant planets are embedded, the velocity perturbations do not create strong traffic jams \citep{Pinilla2015}. The presence of traffic jams is therefore linked to the presence of the ice giants in our simulations, meaning that their impact is non-negligible on the dust substructures. Depending on the masses of these planets, many of the disks observed might therefore present features originating in some traffic jam effect rather than from dust trapped in pressure bumps. This effect has been encountered in the past \citep[e.g.,][]{Rosotti2016} and some studies show that it is possible to disentangle between an over-density of dust caused by a pressure trap or by a traffic jam \citep{Pinilla2017b,Pinilla2017a,Dullemond2018}. Observations at multiple wavelengths is a possible way to distinguish between each mechanism and can therefore help to unveil the number of planets contained in the observed disks.

\begin{figure*}[h]
   \centering   
   \includegraphics[scale=0.265]{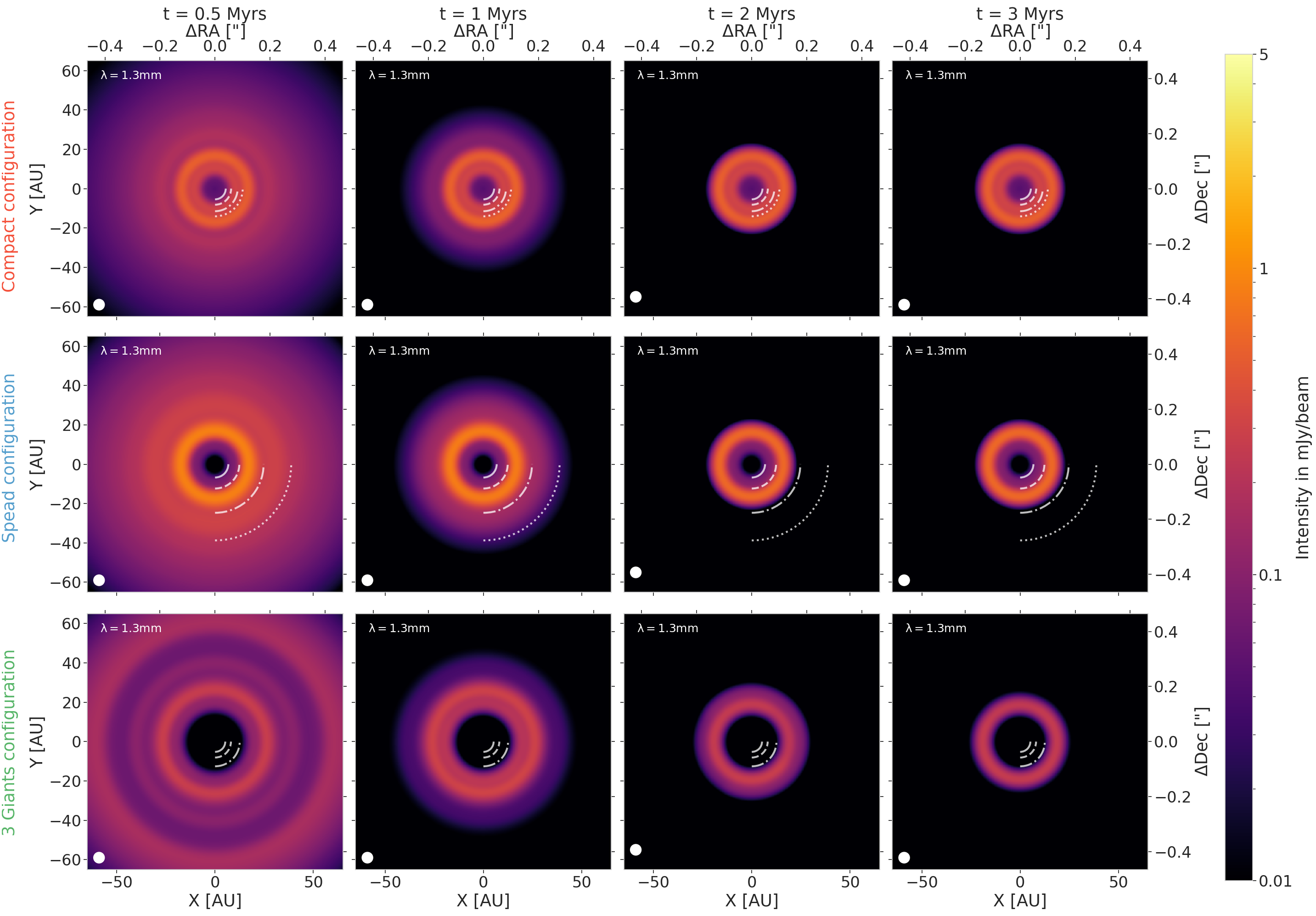}
   \caption{Images at $\lambda =$1.3mm of the different configurations, at low viscosity and low aspect ratio, at different times: 0.5 Myrs, 1 Myr, 2 Myrs, and 3 Myrs, from left to right. The white lines represent the positions of the different planets in each configuration. We assume that the sensitivity of the instrument is limited to fluxes larger than 10 $\mu$Jy/beam. As time evolves, the sizes of the disks shrink due to inward drift, leaving stable rings in the inner disks after 1 Myr.}
   \label{images_difftimes}
\end{figure*}

Another method that can be used to determine if an observed ring is coming from a pressure trap or a traffic jam is to study the CO velocity perturbations \citep{Teague2018,Pinte2018,Pinte2020}. In these studies, the presence of a pressure gradient in the disk can be linked to a change of rotational velocity. Traffic jams, originating from a perturbation of the radial gas velocity, would not influence the rotational velocity profile of the disk. The presence of this "kink" in the CO rotational velocity perturbations could be used to distinguish between a traffic jam or a pressure bump \citep{Izquierdo2021}.

\subsection{Impact of time evolution}
\label{section_discussion_timeevol}

With this project, we made the choice to implement dust growth with TWO-POP-PY at the expense of a 2D or 3D evolution model, allowing us to study the time evolution of dust growth. As \cite{Drazkowska2019} show in their 2D study, including dust coagulation has a non-negligible impact on the dust distribution. In our case, we implemented dust growth by assuming that the gas is fixed during dust growth (here evolved for 1 Myr). This assumption makes a good approximation at low viscosity for two reasons: the first one is that the viscous timescale is way larger than 1 Myr, meaning that the disk would remain almost static during this time; on the other hand, migration is slow at this viscosity \citep{Baruteau2014}, meaning that the dynamics of the planets would not strongly influence the gas disk structure. Even if instabilities can be triggered at low viscosities creating azimuthal asymmetries (as in \cite{Zhang2018}, Section 5.1 with a low viscosity Solar System disk), we assume that these asymmetries vanished by the time the planets are fully formed, as supported also by other hydrodynamical simulations \citep[e.g.,][]{Hammer2017,Bergez2020} and shown in Appendix \ref{appendix_2Dgas}.

However, at high viscosity, gas evolution and planet-disk interactions over 1 Myrs start to be non-negligible: the gas is accreted toward the star and planets migrate faster. Migration of planets can alter the dust distributions in the disk \citep[e.g.,][]{Meru2019,Weber2019}, but, as mentioned in Sect. \ref{section_dustevol}, the substructures in the gas (e.g., pressure bumps) are not strong enough to block the inward diffusing dust, preventing the creation of notable features in the disk. Therefore, our setup is a good approximation to estimate how dust is distributed and further studies would be needed to detail the impact of gas evolution.

\begin{figure}[t]
   \centering   
   \includegraphics[scale=0.27]{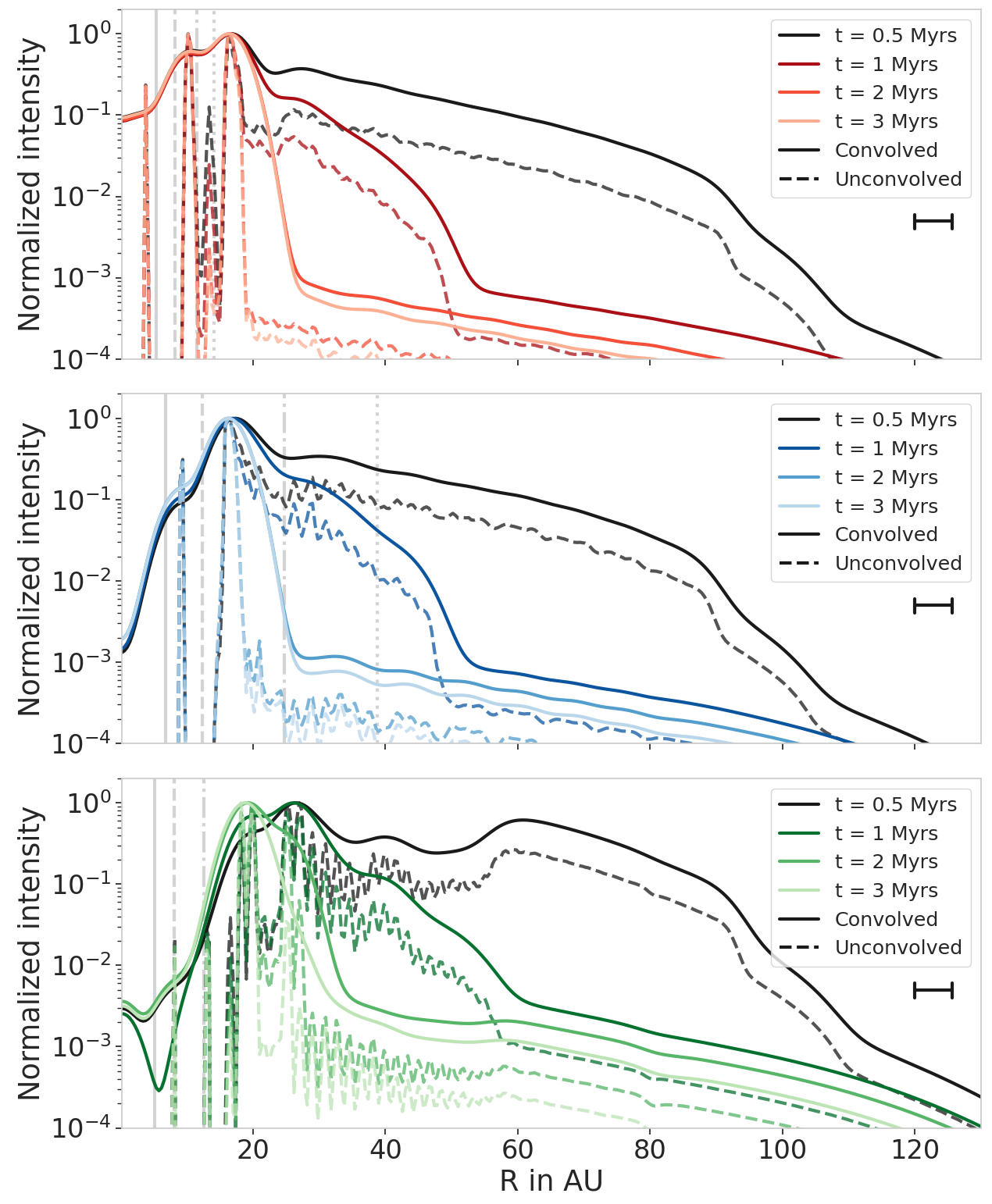}
   \caption{Radial intensity profiles after 0.5 Myr, 1 Myr, 2Myrs, and 3 Myrs of evolution in each configuration with low viscosity and a low aspect ratio. The total size of the disk depends on the time evolution, which determines the dust distribution.}
   \label{profiles_difftimes}
\end{figure}

As mentioned, in this paper, we chose to evolve the dust for 1 Myr during which the gas profile is considered constant. However, as Eqs. \ref{eq_driftlimit} and \ref{eq_growthlimit} show, the millimeter dust disk size will expand with time as dust will grow further and further out in the disk, until drift will deplete the outer regions of the disk and reduce the millimeter disk size. In Figs. \ref{images_difftimes} and \ref{profiles_difftimes} we present the images and their radial intensity profiles at different times: after 0.5 Myrs, 1 Myr (as in Sect. \ref{section_dustevol}), 2 Myrs, and 3 Myrs. After 0.5 Myrs, the dust disk is large as drift only starts to deplete the outer region of the disk, showing more substructures than at later times. In the Spread configuration, the gaps created by Uranus and Neptune are slightly distinguishable before being washed away by drift. In the Three-Giants case, the velocity perturbations induced by the planets create numerous gaps and rings in the outer regions, before being washed at later times by drift as for the other configurations. The presence of numerous substructures in young disks matches the observations of the disk around IRS 63, supposedly younger than 0.5 Myr \cite{SeguraCox2020}. 

After 2 Myrs, the majority of the dust located in the outer disk had time to grow and drift to the inner regions. The only dust remaining is the dust trapped in the pressure bump located at Saturn's gap or at the outer giant gap. \cite{Long2020} also investigated the impact of time evolution of the dust size of the disk: they show that without any dust traps, the millimeter dust size of the disk increases until drift reduces the disk; on the other, in presence of a pressure bump created by a planet and acting like a dust trap, the size of the disk is first dominated by growth and then by the position of the dust trap. 

Furthermore, at low viscosity, some asymmetries in the gas can arise due to instabilities \citep[e.g., Rossby wave instability][]{Lovelace1999,Li2001}: to study these asymmetries, a 2D (at least) analysis has to be done. Future simulations, with sufficient computational power, should then consider both dust growth in multidimensional grids \citep{Drazkowska2019}. In our study, we avoided these asymmetries by simulating the gas disk for sufficient time, letting the possible instabilities dissipate in the disk \citep{Hammer2017}. 

Time evolution also has an impact on the characteristics of the planets. Indeed, we assume that the planets already have their final mass and do not migrate. \cite{Bergez2020} showed that gas accretion can have an impact on the gap depth, specially at low viscosities. This would influence the pressure profile of the disks and therefore the dust distributions and the created substructures. Computing the evolution of gas, planet mass and semimajor axis with dust growth and evolution simultaneously would be ideal but too computationally expensive. In order to investigate what the effect of the planet mass can have on the results, we showed in Sect. \ref{section_diffmassimages} how the images could be different at different stages of planet evolution. The similarities between the images showed us that it is more important to improve the resolution of our observations, as a resolution of 0.02'' or better is needed to disentangle between different planetary systems.



\subsection{Disk dust mass}
\label{section_diskuss_diskmasses}

\begin{figure}[t]
   \centering   
   \includegraphics[scale=0.25]{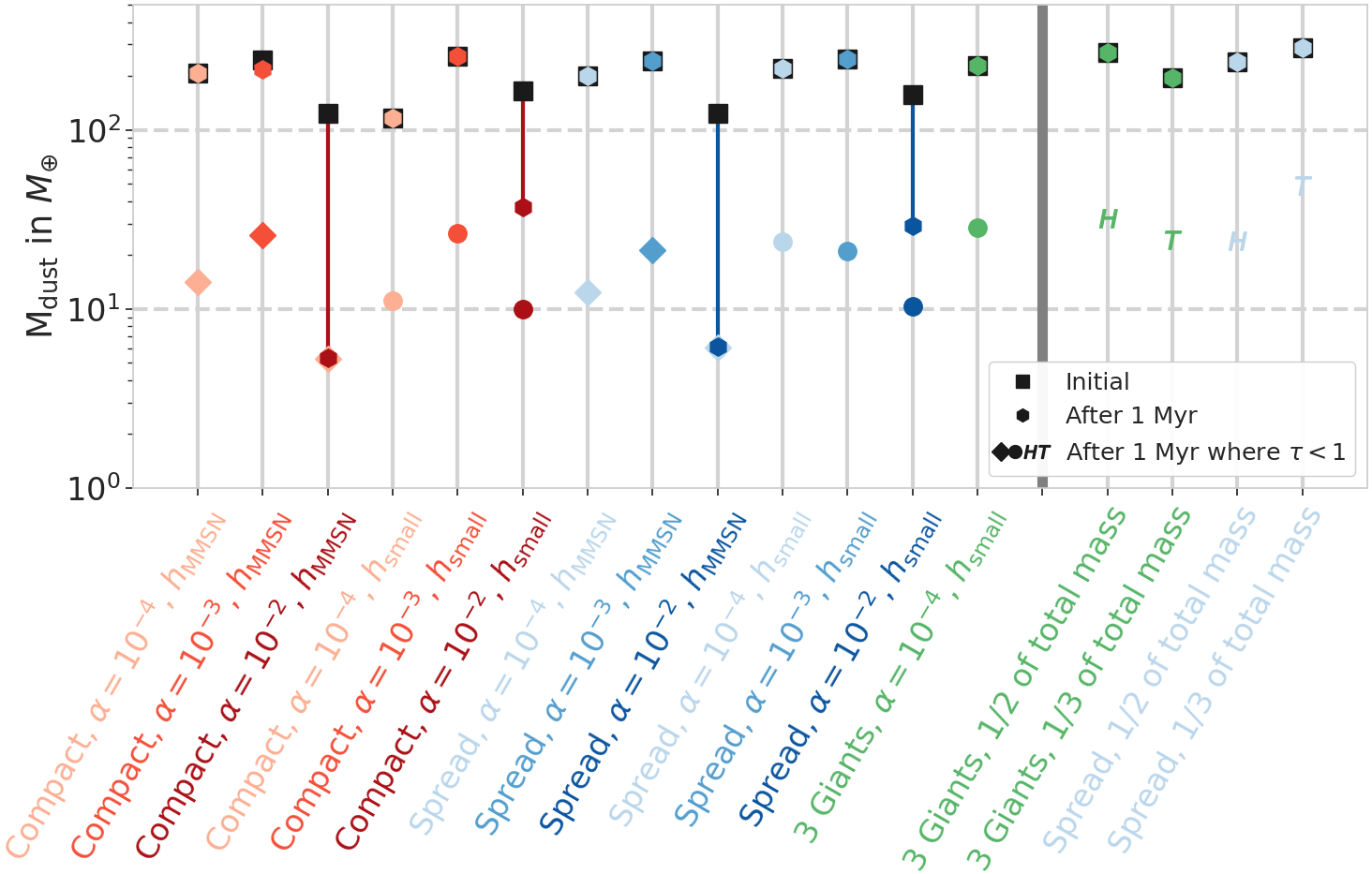}
   \caption{Masses of the dust disks for each configuration and each disk parameter derived from the dust evolution model. The squares correspond to the initial amount of dust mass, and the circles show how much of this mass is left after 1 Myr of dust evolution. The differences in initial dust masses for all disks originate from the assumption that the dust-to-gas ratio is initially 0.01 and the inner disk is depleted in dust by the giant planets (see Sect. \ref{section_dustevol}). In most cases, all the dust is trapped in the disks, except at high viscosity because the planet gaps are not strong enough to prevent the dust from flowing toward the star. To compare to observed dust masses as in Fig.\ref{mobs_moptthin}, we also derive the dust contained in the optically thin regions (represented by the markers corresponding to Fig. \ref{mobs_moptthin}). We see here that the majority of the mass is hidden in the optically thick regions.}
   \label{diff_dustmasses}
\end{figure}

One of the challenges of planet formation is to overcome the too large radial drift speed of dust. Cosmochemical studies show that planetesimal formation happened at all times during the Solar System disk age \citep[e.g.,][]{Connelly2012}. Therefore, if the dust drifts too rapidly during the disk lifetime, then there is not enough material to to ensure a continuous planetesimal formation as in the Solar System. One way to prevent fast inward drift that empties the disk is to trap the dust \citep{Pinilla2012}, as showed in Sect. \ref{section_dustevol}. We investigate here how much dust is actually remaining in our disks as a function of disk parameters and planet configurations. In Fig.\ref{diff_dustmasses} we show the total dust mass contained in each of our disks at two different times: initially and after 1 Myr of evolution. 

As expected from Sect. \ref{section_dustevol}, the high viscosity disks ($\alpha = 10^{-2}$) do not present strong dust traps and the high diffusion allows the dust to drift toward the star. Therefore, after 1 Myr, they lost a significant amount of dust. On the other hand, all the disks with lower viscosities ($\alpha = 10^{-3}$ and $10^{-4}$) trap the dust efficiently. Even if the dust is distributed differently as a function of time (see Figs. \ref{images_difftimes} and \ref{profiles_difftimes} from the previous section), the amount of trapped dust available to form the small bodies of our Solar System stays constant (from the time the giant planets form and assuming that the inner disk is already depleted in dust).

In all the disks with efficient dust trapping, the total amount of dust is above 100 $\rm M_\oplus$. This can be compared to the solid masses needed to form the small bodies of the Solar System. The solid budget needed to trigger the Nice instability is of at least 20 $\rm M_\oplus$ \citep{Nesvorny2012,Nesvorny2013}. As the mass contained in the asteroid belt is negligible ($5\times10^{-4}\rm M_\oplus$, \cite{Kresak1977}), our dust disks have enough material to form the small bodies of the Solar System even if the formation of planetesimal from pebbles is not 100$\%$ efficient. Moreover, having the majority of the dust located in the outer disk (i.e., outside Saturn's location) is in agreement with \cite{Izidoro2021} where the authors show that the inner Solar System, after formation of some planetesimals, should be depleted in pebbles to explain the formation of the current terrestrial planets. This also clearly indicates that the viscosity of the gas in the protoplanetary disk must have been low, because at high viscosity even Jupiter is not able to block inward drifting pebbles, which is required by cosmo-chemical studies \citep[e.g.,][]{Kruijer2017,Weber2018}.

On the other hand, dust masses derived by observations are on the order of a few Earth masses to a few tens of Earth masses \citep{Andrews2013,Ansdell2016}. In order to understand the origin of this discrepancy between the dust masses in our simulations and the one derived by observations, we derive the dust masses from our disk images using the same methods as in the observations and compare them to the actual dust mass of our simulations. Assuming an optically thin disk, the observed dust mass $M_{d,obs}$ is given by \citep{Hildebrand1983}\begin{equation}
    M_{d,obs} = \frac{F_\nu \, d^2}{\kappa_\nu^{abs} B_\nu(T_d)}
    \label{eq_mdustobs}
,\end{equation}
where $F_\nu$ is the total integrated flux density of the disk, $d$ the disk's distance, $\kappa_\nu^{abs}$ the absorption opacity at the observed wavelength (here $\lambda = 1.3$ mm) and $B_\nu(T_d)$ the Planck function at temperature $T_d$. 

In general, observers use the assumptions made by \cite{Andrews2013} regarding the opacity and dust temperature. The opacity is assumed to be $\kappa_\nu^{abs} = 2.3 \;\rm cm^2/g$, following \cite{Beckwith1990}. The averaged temperature is taken as $T_d = 20K$. This value was derived from the observations of Taurus disks by \cite{Andrews2005}, using Eq. \ref{eq_mdustobs}: the authors assumed a simple disk model in order to determine the disk dust mass and derived the disk average temperature assuming the same opacity as mentioned above. To be consistent with our simulations, we derive the observed masses using the absorption opacity and average temperature from the RADMC3D outputs and compare the resulting masses to the ones obtained by following \cite{Andrews2013}. Assuming that mostly millimeter grains contribute to the opacity at this wavelength \citep{Dullemond2018}, we use the absorption opacity derived from OpTool following the assumptions made in Sect. \ref{section_dustevol_setup}, resulting in $\kappa_\nu^{abs} = 2.04 \;\rm cm^2/g$ for a grain size of $a = 0.1 \rm \; cm$ at $\lambda = 1.3\rm \; mm$. The dust disk temperatures and masses can be found in Table \ref{table_dust_masses}.

\begin{table*}[t]            
\centering                          
\caption{Dust masses calculated from the total integrated flux as in Eq. \ref{eq_mdustobs}.}
\begin{tabular}{l c c | c c  c |c c}        
\hline
\hline             
 & & & \multicolumn{3}{|c|}{This paper} & \multicolumn{2}{c}{\cite{Andrews2013} setup} \\
 (1) & (2) & (3) & (4) & (5) & (6) & (7) & (8) \\
Configuration & $F_\nu$ & $M_{d,tot}$ & $T_d$ & $M_{d,\tau < 1}^{tot}$ &  $M_{d,obs}^{\lambda = 1.3 \rm \; mm}$ & $M_{d,\tau < 1}^{tot}$ & $M_{d,obs}^{\lambda = 1.3 \rm \; mm}$ \\    
 & (mJy) & ($M_\oplus$) & (K) & ($M_\oplus$) & ($M_\oplus$) & ($M_\oplus$) & ($M_\oplus$) \\
 
\hline                   
\textcolor{Ca4}{Compact, $\alpha = 10^{-4}$, MMSN $h$} & 48.9 & 206.8 & 41.8 & 14.1 & 12.6 & 13.1 & 27.1 \\
\textcolor{Ca3}{Compact, $\alpha = 10^{-3}$, MMSN $h$} & 48.3 & 216.7 & 39.6 & 26.0 & 13.2 & 23.5 & 26.8\\
\textcolor{Ca2}{Compact, $\alpha = 10^{-2}$, MMSN $h$} & 7.8 & 5.33 & 45.8 & 5.3 & 1.8  & 5.3  & 4.3\\
\textcolor{Ca4}{Compact, $\alpha = 10^{-4}$, low $h$} & 19.8 & 115.9 & 46.8 & 11.0 & 4.5 & 11.0 & 11.0\\
\textcolor{Ca3}{Compact, $\alpha = 10^{-3}$, low $h$} & 48.0 & 258.9 & 43.2 & 26.4 & 11.9 & 24.4 & 26.6 \\
\textcolor{Ca2}{Compact, $\alpha = 10^{-2}$, low $h$} & 30.1 & 37.1  & 45.4 & 9.9 & 7.1 & 9.5 & 16.7 \\
 & & & & & & & \\
\textcolor{Sa4}{Spread, $\alpha = 10^{-4}$, MMSN $h$} & 41.9 & 199.4 & 43.5 & 12.4 & 10.3 & 12.3 & 23.3 \\
\textcolor{Sa3}{Spread, $\alpha = 10^{-3}$, MMSN $h$} & 54.9 & 243.3 & 40.1 & 21.2 & 14.8 & 20.5 & 30.5 \\
\textcolor{Sa2}{Spread, $\alpha = 10^{-2}$, MMSN $h$} &  8.9 &   6.1 & 45.6 & 6.1 & 2.1 & 6.1 & 5.0 \\
\textcolor{Sa4}{Spread, $\alpha = 10^{-4}$, low $h$} & 28.8 & 220.7  & 45.8 & 23.8 & 6.7 & 22.7 & 16.0\\
\textcolor{Sa3}{Spread, $\alpha = 10^{-3}$, low $h$} & 51.2 & 248.7 & 45.2 & 21.0 & 12.1 & 19.9 & 28.4 \\
\textcolor{Sa2}{Spread, $\alpha = 10^{-2}$, low $h$} & 31.6 & 29.0  & 45.6 & 10.3 & 7.4 & 10.1 & 17.6 \\
& & & & & & & \\
\textcolor{Ga4}{3 Giants, $\alpha = 10^{-4}$, low $h$} & 18.8 & 229.2 & 51.0 & 28.4 & 3.9 & 26.9 & 10.5 \\
\textcolor{Ga4}{3 Giants, 1/2 of total planet mass}    & 31.1 & 269.5 & 47.9 & 32.1 & 6.9 & 30.2 & 17.3\\
\textcolor{Ga4}{3 Giants, 1/3 of total planet mass}    & 36.1 & 196.6 & 46.8 & 24.2 & 8.2 & 22.6 & 20.1 \\
\textcolor{Sa4}{Spread  , 1/2 of total planet mass}    & 31.3 & 240.2 & 45.6 & 24.0 & 7.3 & 23.7 & 17.4 \\
\textcolor{Sa4}{Spread  , 1/3 of total planet mass}    & 52.5 & 288.6 & 44.9 & 49.1 & 12.5 & 47.8 & 29.2 \\

\hline
    
\end{tabular}

\tablefoot{Columns are: (1) Planet and disk configuration. (2) Total integrated flux density in mJy. (3) Total dust mass from the dust evolution model after 1 Myr of evolution. (4) Average temperature from the RADMC3D outputs. (5) Actual mass contained in the optically thin region, assuming $\tau_\nu(r) = \Sigma(r)\times\kappa_\nu^{abs}$. (6) Mass derived from observations at $\lambda = 1.3\rm \; mm$, derived by Eq. \ref{eq_mdustobs}. (7) Actual mass contained in the optically thin region, assuming $\kappa_\nu^{abs} = 2.3 \rm \; cm^2/g$ as in \cite{Andrews2013}. (8) Mass derived from observations at $\lambda = 1.3\rm \; mm$, derived by Eq. \ref{eq_mdustobs} assuming $T_d = 20K$ as in \cite{Andrews2013}. On average, our disk temperature is around $45K$, which is more than twice the usual dust disk temperature used in observational studies. Comparing the total mass (3) to the mass contained in the optically thin regions (5), we show that the majority of the mass is hidden in optically thick regions, like, for example, our dust rings. } 
\label{table_dust_masses}
\end{table*}

Eq. \ref{eq_mdustobs} relies on the assumption that the disk is optically thin. However, from the dust evolution models in Sect. \ref{section_dustevol}, we know that the dust is trapped in relatively dense rings that can become optically thick. As some significant amount of mass can be hidden in optically thick regions, we compare the observed dust mass $M_{d,obs}$ to the dust mass contained in the optically thin regions. In order to do so, assuming $\tau_\nu(r) = \Sigma_d(r)\times \kappa_\nu^{abs}$, the comparison is made with the amount of dust contained in the regions where $\tau < 1$. 

By comparing the total amount of dust to the mass contained in the optically thin regions (Cols. 3 and 5 or circles and diamonds in Fig.\ref{diff_dustmasses}), we show that the majority of the mass is hidden in the optically thick regions. This results in masses $M_{d,\tau_\nu < 1}$ that can be more than ten times lower than the actual total mass. Therefore, the presence of optically thick rings in disk images can hide a large amount of dust.

\begin{figure}[t]
   \centering   
   \includegraphics[scale=0.19]{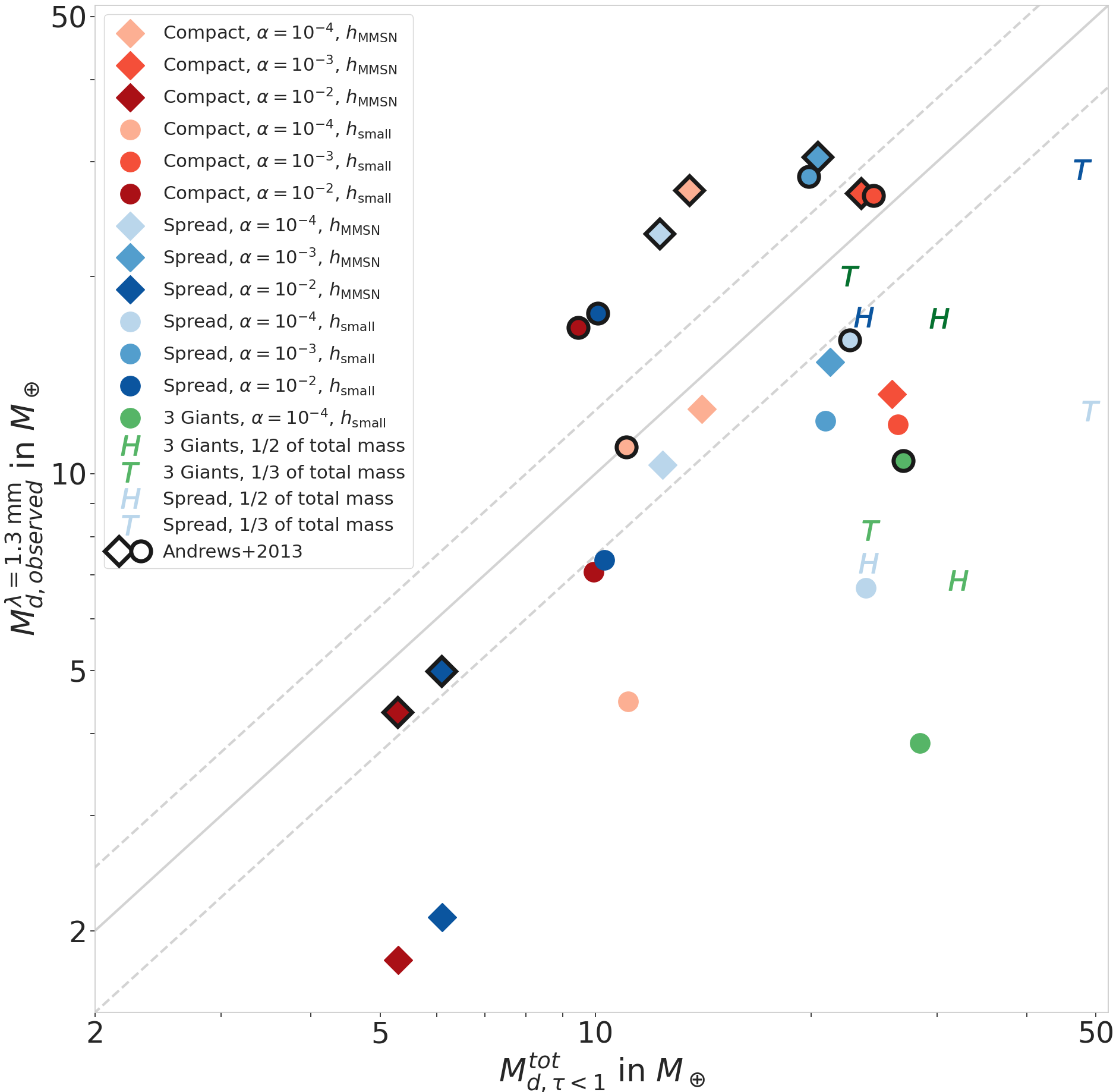}
   \caption{Observed dust masses for each configuration (colors) and each disk parameter (markers). A black outline surrounds the masses derived using the \cite{Andrews2013} assumptions for the disk's opacity and temperature. $M_{d,\tau_\nu < 1}^{tot}$ corresponds to Cols. 5 and 7 (see also Fig.\ref{diff_dustmasses}) and $M_{d,obs}^{\lambda = 1.3 \rm \;mm}$ to Cols. 6 and 8 in Table \ref{table_dust_masses}. The solid gray line represents the masses for which the observations match the masses from the simulations. The top (bottom) dashed line shows the disks for which the observations overestimate (underestimate) the actual mass contained in the optically thin part by 25$\%$.}
   \label{mobs_moptthin}
\end{figure}

We show in Fig.\ref{mobs_moptthin} the difference between the masses derived from Eq. \ref{eq_mdustobs}, $M_{d,obs}^{\lambda = 1.3 \rm \;mm}$ and the masses contained in the optically thin regions of our disks, $M_{d,\tau_\nu < 1}^{tot}$. The diamonds and hexagons represent the different disk parameters where the colors match the planet configuration (Table \ref{table:1}). Markers surrounded by a black outline are the masses assuming the \cite{Andrews2013} setup. We clearly see that the masses are under-estimated when using the actual dust temperature average, whereas a colder temperature (as in \cite{Andrews2013}) tends to over-estimate it.

The underestimation of the observed dust mass derived with Eq.\ref{eq_mdustobs} with self-consistent opacities and disk temperatures originates from the calculation of $M_{d,\tau_\nu < 1}^{tot}$. As this mass comes from our simulations, it contains all grain sizes present in the disk. However, as $M_{d,obs}^{\lambda = 1.3 \rm \;mm}$ is based on $F_\nu$ (Eq. \ref{eq_mdustobs}), it represents only the grains emitting consequently at the observed wavelength. In our case, the image at $\lambda = 1.3 \rm \;mm$ is dominated by millimeter grains, meaning that the mass contained in the smaller grains is not probed here, leading to a less massive dust disk. As this mass underestimation is therefore expected, this shows that the $T_d = 20K$ assumption is leading to an unrealistic overestimation of the optically thin dust masses from observations. From our study, an average temperature of 45 K (see Table \ref{table_dust_masses}) might give more reasonable results; however, further complete studies can help in improving the estimation of $T_d$. 

In conclusion, observations might completely under-estimate the total amount of dust mass contained in disks due to optically thick regions, as it was shown in other studies \citep{Dullemond2018}. Improving our understanding on opacities and disks temperatures are crucial to unveil the mystery around the amount of material available for planet formation. 

\section{Conclusions}
\label{section_conclusions}

In this paper we derived images at $\lambda = 1.3 \,mm$ of different planetary system configurations representing the potential Solar System protoplanetary disk. We also derived the images of a giant system composed of three planets of 1 Jupiter mass each, representing a potential initial state for scattering events to happen and produce eccentric planets that match the radial velocity observations \citep{Juric2008,Raymond2009a,Sotiriadis2017,Bitsch2020}. Using 2D hydrodynamical simulations we determined the gas disk profile in the presence of four (or three) giant planets. This profile was then used as an input for a dust evolution model. After 1 Myr of dust evolution, the resulting dust distributions were used to compute synthetic images of these different disks. Our main conclusions are:

\begin{itemize}
    \item The dust distributions show that the perturbations created by multiple planets in one disk can lead to substructures that are not directly linked to the positions of the planets. These features are created by traffic jams in the disk, revealing the importance of the gas radial motion in the case of multiple giant planets. Considering that a single planet can also create multiple gaps and rings in a low viscosity disk by perturbing the gas surface density \citep{Dong2018,Bae2018}, this complicates the relation between the number of features created by single or multiple planets and the actual number of perturbers. Our study thus highlights that not all individual gaps and rings are caused by individual planets perturbing the pressure profile of the disks, complicating the link between protoplanetary disk observations and exoplanets.

    \item By comparing the synthetic images obtained to known observed disks, we showed that the disk phase can be used to derive robust constraints on planet formation scenarios. The presence of bright substructures located at large radii in the DSHARP survey can be explained by the large size, mass, and brightness of these disks. Here, we show that planet formation occurring in smaller disks can easily be missed at low resolutions in the observations (i.e., with a beam larger than $0.04"\times0.04"$). One way to improve our understanding of planet formation is thus to observe small protoplanetary disks (i.e., of a few tens of AU) at high resolution to probe the formation environments of different planetary populations.
    
    \item The Three-Giants configuration, representing a future system that could experience scattering events after the disk phase, only presents substructures within 40 AU. While \cite{Lodato2019} speculate that the bright rings observed by DSHARP can be explained by the presence of fast migrating giant planets matching the radial distribution of eccentric planets observed by radial velocity, \cite{Ndugu2019} show that this requires a migration at high viscosity, which is contrary to the recent derivation of disk viscosity. Our study here shows that a giant planet system that is susceptible to scatter later during its formation would not produce bright rings in the outer regions during its gas disk phase.
    
    \item At high viscosity, too much dust diffuses through the gaps generated by Jupiter and Saturn, inconsistent with terrestrial planet formation \citep[e.g.,][]{Izidoro2021} and cosmo-chemical evidence \citep[e.g.,][]{Kruijer2017}. At low viscosity, dust can be retained in a pressure trap exterior to the giant planets, generating large optically thick dust pileups. Self-consistently constraining the dust mass of the disks observationally revealed that the observationally inferred dust mass can be a factor of ten below the real dust mass in optically thick rings in our simulations. Moreover, improving dust temperature estimates can highly improve the estimation of dust from the observations.
    
\end{itemize}

This study shows the importance of resolution in observations for our understanding of planet formation. For example, in the Compact configuration (Figs. \ref{profile_hMMSN}, \ref{profile_hsmall}, \ref{image_1.3mm_hMMSN}, and \ref{image_1.3mm_h4}), the features created by the four giant planets were smeared out by the beam of the instrument, making it impossible to determine how many planets are located in this disk. If future surveys focus on very high resolution observations of smaller protoplanetary disks, then it will be possible to distinguish the conditions needed for giant planets to form in the outer or inner regions of the disk. As discussed in Sect. \ref{section_comparison_ALMA}, an interferometer such as ALMA already has the power to produce images with a high enough resolution. Such observations should be combined with further studies that model the disk structures in the presence of multiple planets. Finally, in order to improve our understanding of the origin of the dust substructures (traffic jams or pressure bumps as discussed in Sect. \ref{section_diskuss_trafficjams}), multiwavelength imaging will help us determine how many planets are trapped in disks, as well as help us determine the optical properties of the dust. This last point is important for deriving how much mass is available in disks for planet formation.

\begin{acknowledgements}
      C. Bergez-Casalou and B. Bitsch thank the European Research Council (ERC Starting Grant 757448-PAMDORA) for their financial support. C. Bergez-Casalou thank K. Dullemond for his help in managing RADMC3D and T. Birnstiel, C. Lenz and M. Garate for their help in managing TWO-POP-PY and F. Cantalloube and J. Milli for their interesting discussions on observations at different wavelengths. P. Pinilla and N.T. Kurtovic are also thankful for the support provided by the Alexander von Humboldt Foundation in the framework of the Sofja Kovalevskaja Award endowed by the Federal Ministry of Education and Research. The authors thank the referee for their interesting remarks and questions that helped improve this paper.
\end{acknowledgements}

\bibliographystyle{aa}
\bibliography{biblio}

\begin{appendix}
\section{Gas hydrodynamical profiles}
\label{appendix_2Dgas}

\begin{figure*}[t]
   \centering   
   \includegraphics[scale=0.33]{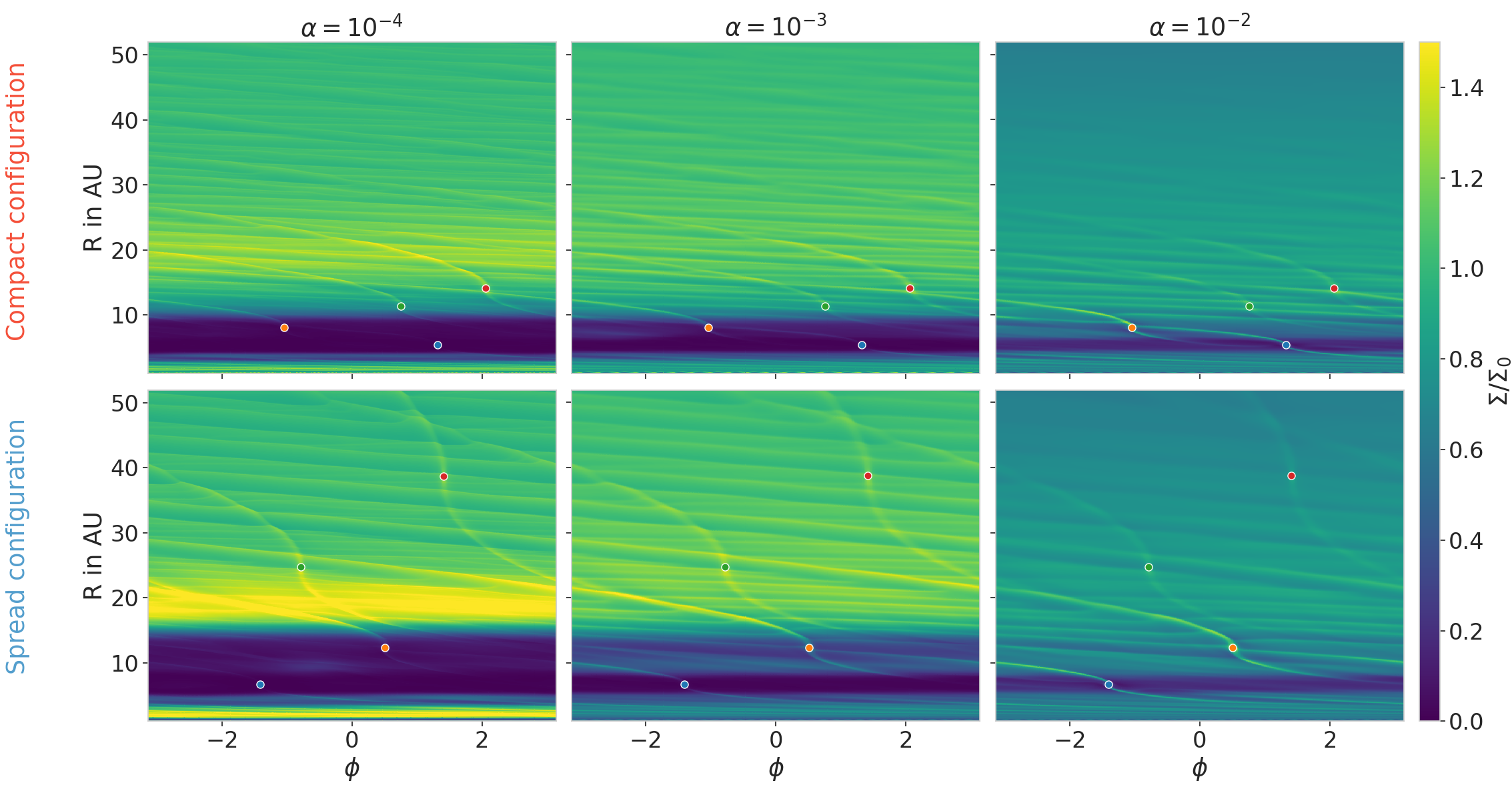}
   \caption{Perturbed gas surface densities ($\Sigma/\Sigma_{0}$) for an MMSN-like aspect ratio at t = 12 500 orbits of the inner planet, in the Compact (first row) and Spread (second row) configurations. The positions of the planets are marked by dots in each panel (blue corresponds to Jupiter, orange to Saturn, green to Uranus, and red to Neptune). The disks can be considered axisymmetric, which is important for the dust evolution model that takes the 1D radial gas profile as an input.}
   \label{dens2D_hMMSN}
\end{figure*}

\begin{figure*}[t]
   \centering   
   \includegraphics[scale=0.33]{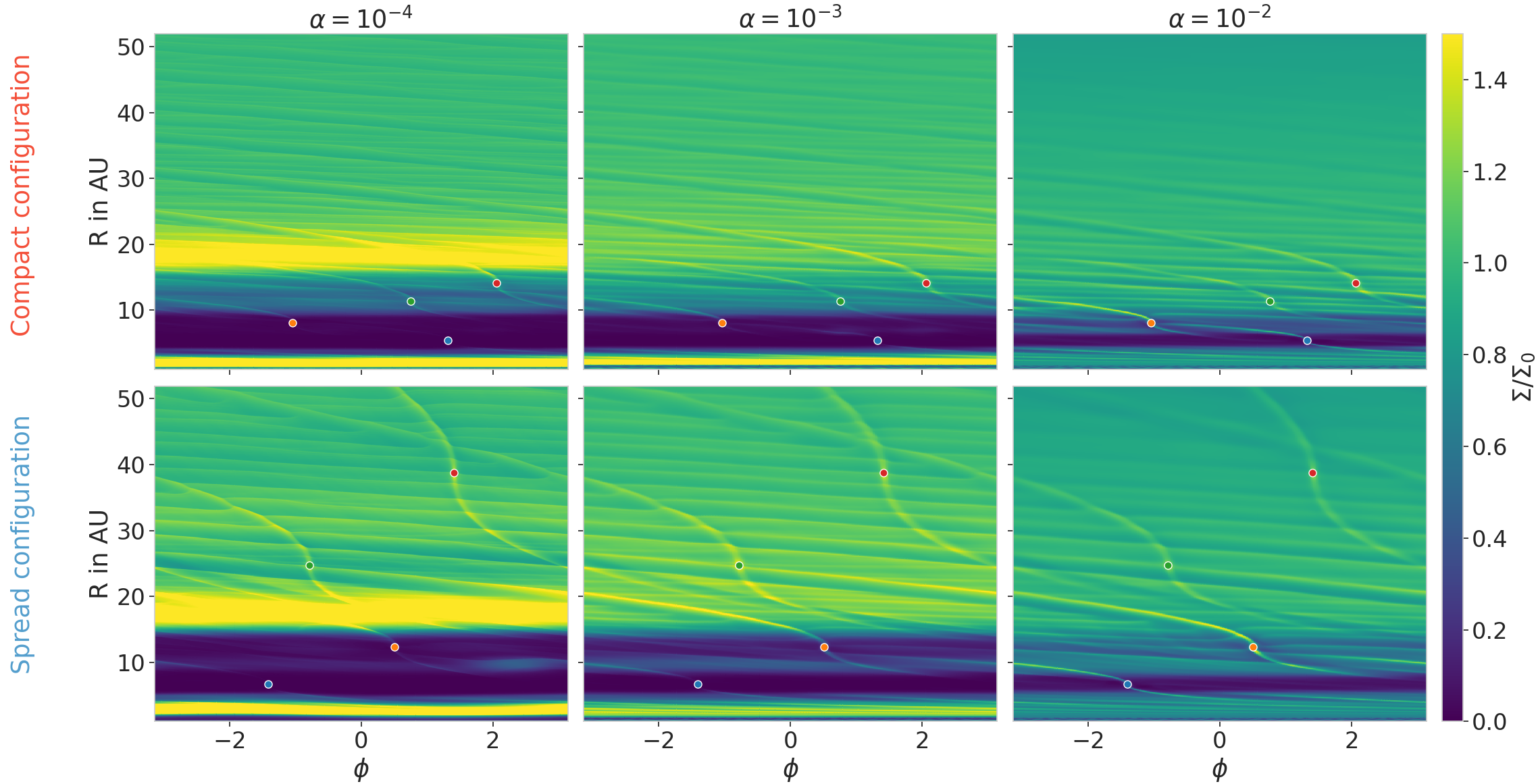}
   \caption{Same as \ref{dens2D_h4} but for a smaller aspect ratio.}
   \label{dens2D_h4}
\end{figure*}

\begin{figure*}[t]
   \centering   
   \includegraphics[scale=0.32]{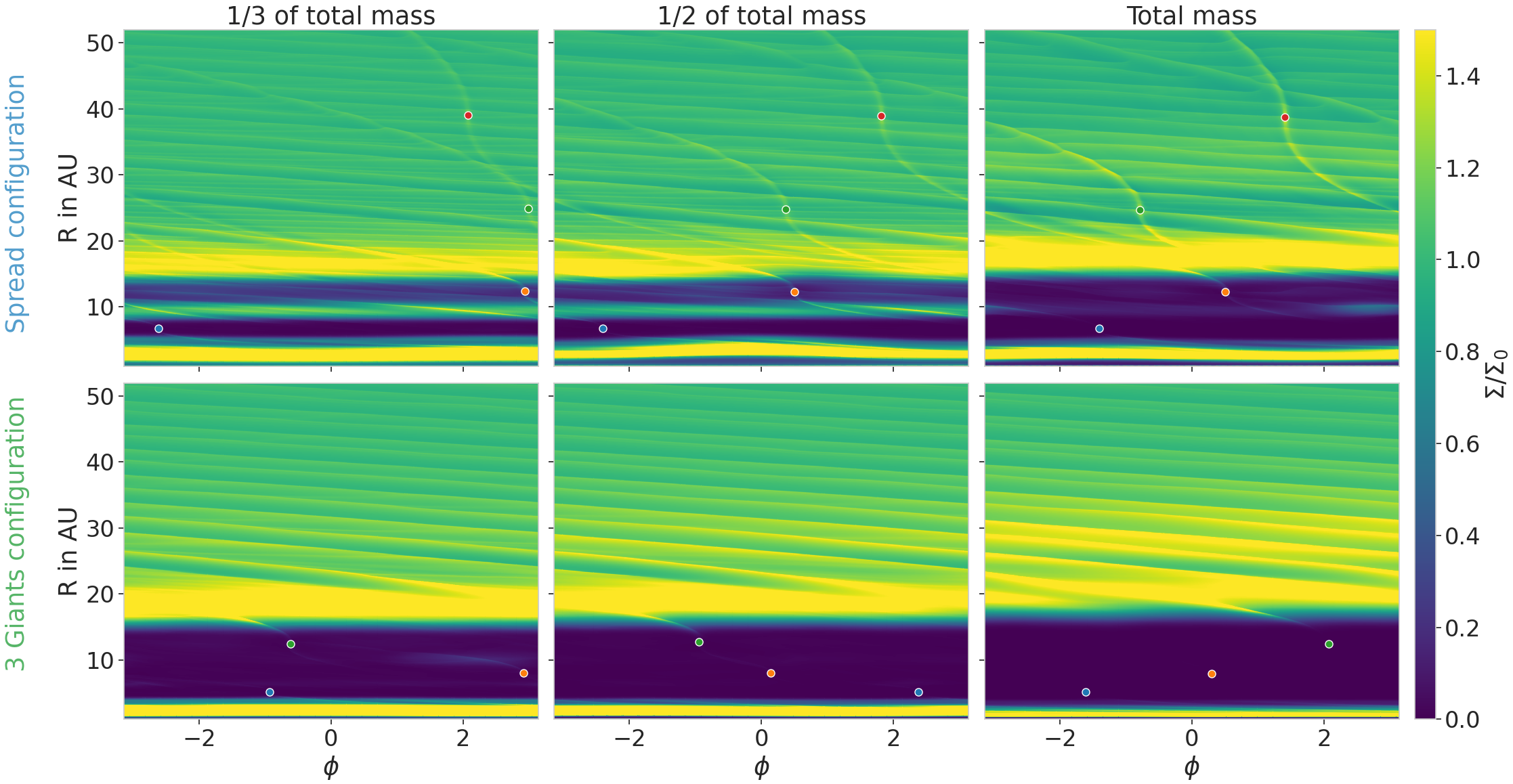}
   \caption{Perturbed gas surface densities ($\Sigma/\Sigma_{0}$) at t = 12 500 orbits of the inner planet, in the Spread (first row) and Compact (second row) configurations. The disks have a small aspect ratio and low viscosity ($\alpha = 10^{-4}$). The masses of the planets are reduced by a factor of two-thirds (left panels) and one-half (middle panels). They can be compared to the total mass configuration in the right panels. As in Fig. \ref{dens2D_hMMSN}, the positions of the planets are marked by dots in each panel and the disks can be considered axisymmetric.}
   \label{dens2D_masscomp}
\end{figure*}

As discussed in Sect. \ref{section_hydro_setup}, we present in this appendix the outputs of the 2D hydrodynamical setups. In Figs. \ref{dens2D_hMMSN}, \ref{dens2D_h4}, and \ref{dens2D_masscomp} we show the perturbed surface densities of the 2D grids of the disks, for the different configurations and disk parameters. Each row represents a configuration, and each column represents a different $\alpha$ viscosity, ranging from $10^{-4}$ to $10^{-2}$ from left to right. The two first figures represent the two different aspect ratios investigated: an MMSN-like aspect ratio in is shown in Fig. \ref{dens2D_hMMSN} and a smaller aspect ratio, as described in Sect. \ref{section_hydro_setup}, in Fig. \ref{dens2D_h4}). In Fig. \ref{dens2D_masscomp} we present the perturbed surface densities for the different masses investigated in the Spread and Three-Giants configurations. In each of the panels of these three figures we see that the gas disk is axisymmetric after t = 12 500 orbits. The vortices triggered by some instabilities or planet growth that could form at low viscosity at the edges of the giant gaps have time to vanish \citep{Hammer2017,Bergez2020}, meaning that we can take the azimuthal average needed as inputs for TWO-POP-PY.

\begin{figure*}[t]
   \centering   
   \includegraphics[scale=0.33]{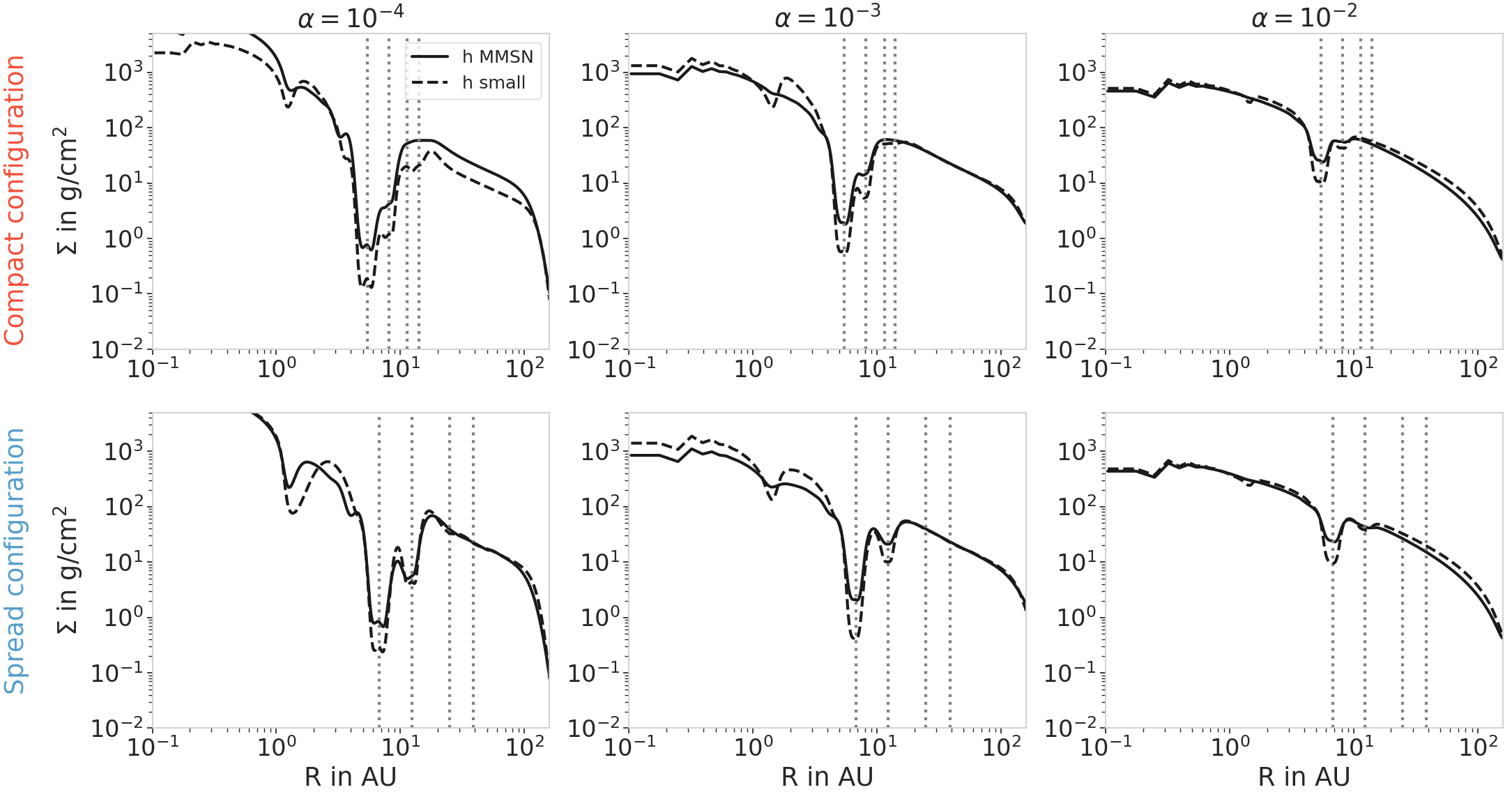}
   \caption{Time- and azimuthal-averaged gas surface density profiles for each aspect ratio. Vertical dotted lines represent the positions of each planet. Jupiter and Saturn are the only ones creating substructures in the disks, except in the Compact configuration with a low $\alpha$ and small aspect ratio, where Uranus and Neptune create a small gap and an over-density outside of Neptune's gap.}
   \label{dens1D_hMMSNh4}
\end{figure*}

\begin{figure*}[t]
   \centering   
   \includegraphics[scale=0.33]{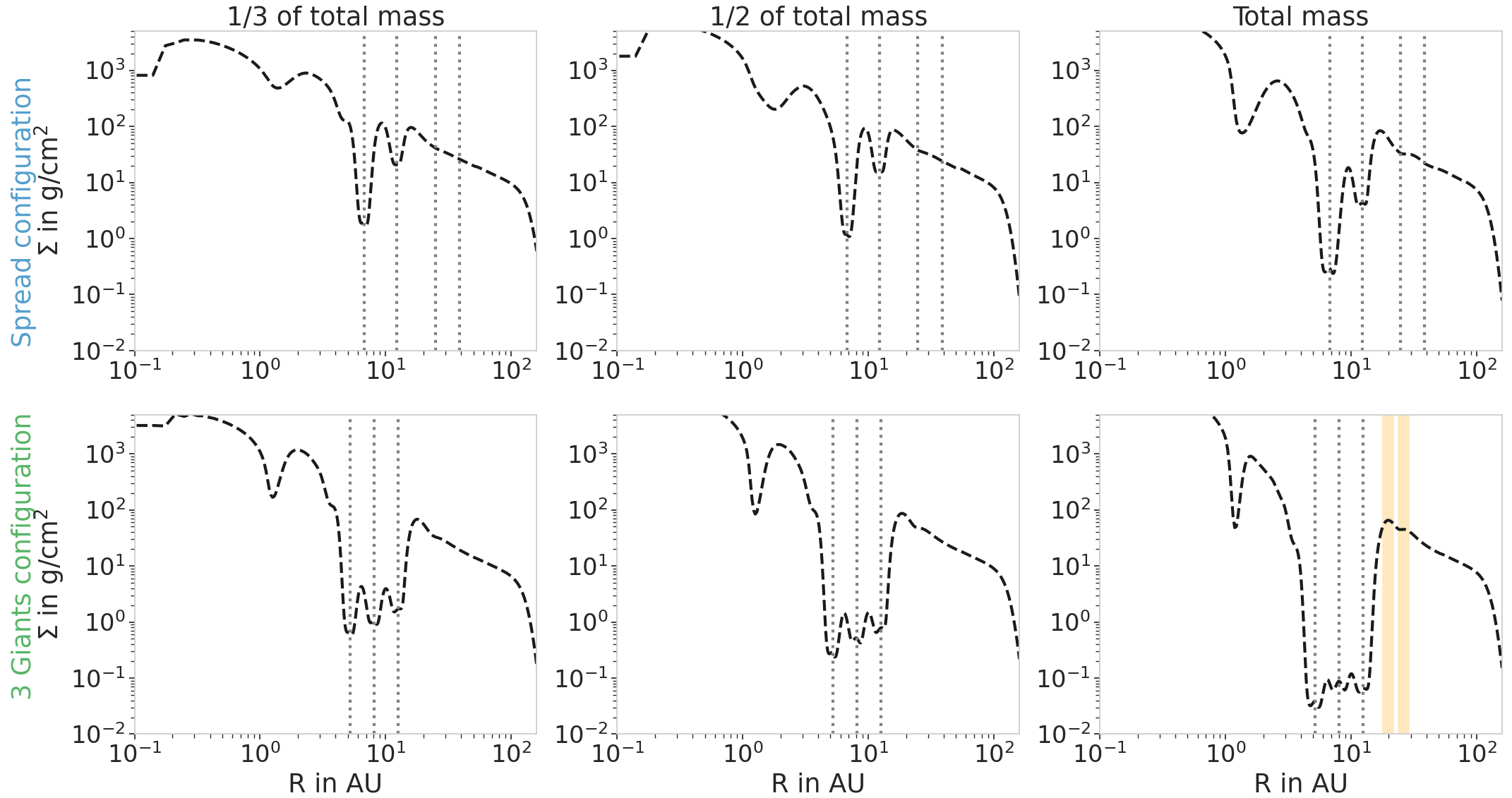}
   \caption{Time- and azimuthal-averaged gas surface density profiles for the different configurations. Vertical dotted lines represent the positions of each planet. As expected, more massive planets create deeper gaps. In the Three-Giants case, the giants are close enough to one another to always create a common gap. The masses of the planets then dictate how deep the gap is and how much gas is present in the gap between them. The two orange vertical lines show the positions of the rings seen in the synthetic millimeter images (Fig. \ref{profile_diffmasses}). We see that they do not correspond to strong features in the gas disk and are located far from the giants' orbits.}
   \label{dens1D_diffmass}
\end{figure*}

In Figs. \ref{dens1D_hMMSNh4} and \ref{dens1D_diffmass} we present the azimuthal and time average gas profiles used as inputs for the dust evolution model (Sect. \ref{section_dustevol}). The profiles are time-averaged over 2 500 orbits. For each configuration and each viscosity, the profiles at each aspect ratio are plotted in the same panel: the MMSN-like aspect ratio is presented in solid line while the smaller aspect ratio is shown in dashed lines. Each planet's orbit is represented by a vertical dotted gray line.

These 2D surface densities show the importance of the viscosity. In the Solar System configurations (Figs. \ref{dens2D_hMMSN} and \ref{dens2D_h4}), Jupiter and Saturn create a common gap at low viscosity whereas only Jupiter is able to start to form a gap at high viscosity. Depending on the planet configuration, at $\alpha = 10^{-3}$, the two inner giants create different features: when they create a common gap in the Compact configuration, some gas is accumulated in between Jupiter and Saturn in the Spread configuration. 

In the Compact configuration, at low viscosity and for both aspect ratios, Uranus and Neptune are massive enough to start creating a pileup of gas outside of Neptune's orbit. It is particularly visible with the small aspect ratio and in the 1D profiles (see Fig. \ref{dens1D_hMMSNh4}). When we compare these two panels to the two corresponding panels for the Spread configuration, we notice that Uranus and Neptune barely have an effect on the gas disk.

Regarding the planets of different mass and the Three-Giants configuration (Figs. \ref{dens2D_masscomp} and \ref{dens1D_diffmass}), we see that the Three-Giants configuration always creates a deep common gap as the planets are close to one another. However, in the Spread configuration the amount of gas present between Jupiter and Saturn clearly create two different gaps when the planets have reduced masses. 

\begin{figure*}[t]
   \centering   
   \includegraphics[scale=0.33]{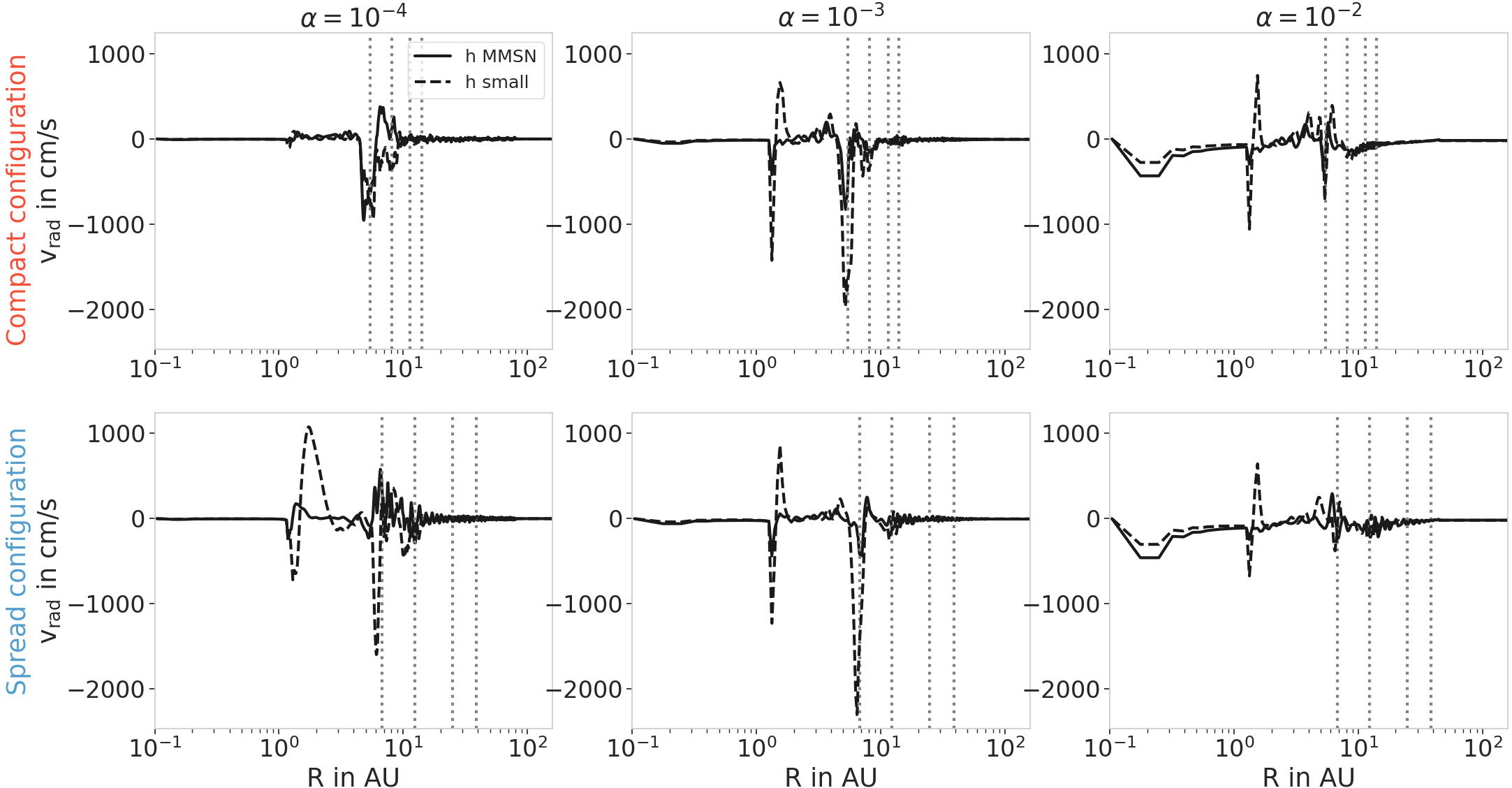}
   \caption{Time- and azimuthal-averaged gas radial velocity profiles for the different configurations. Vertical dotted lines represent the positions of each planet. Multiple planets highly perturb the gas velocities, having an important impact on the dust distributions (see Sect. \ref{section_dustevol}).}
   \label{vrad1D_hMMSNh4}
\end{figure*}

\begin{figure*}[t]
   \centering   
   \includegraphics[scale=0.33]{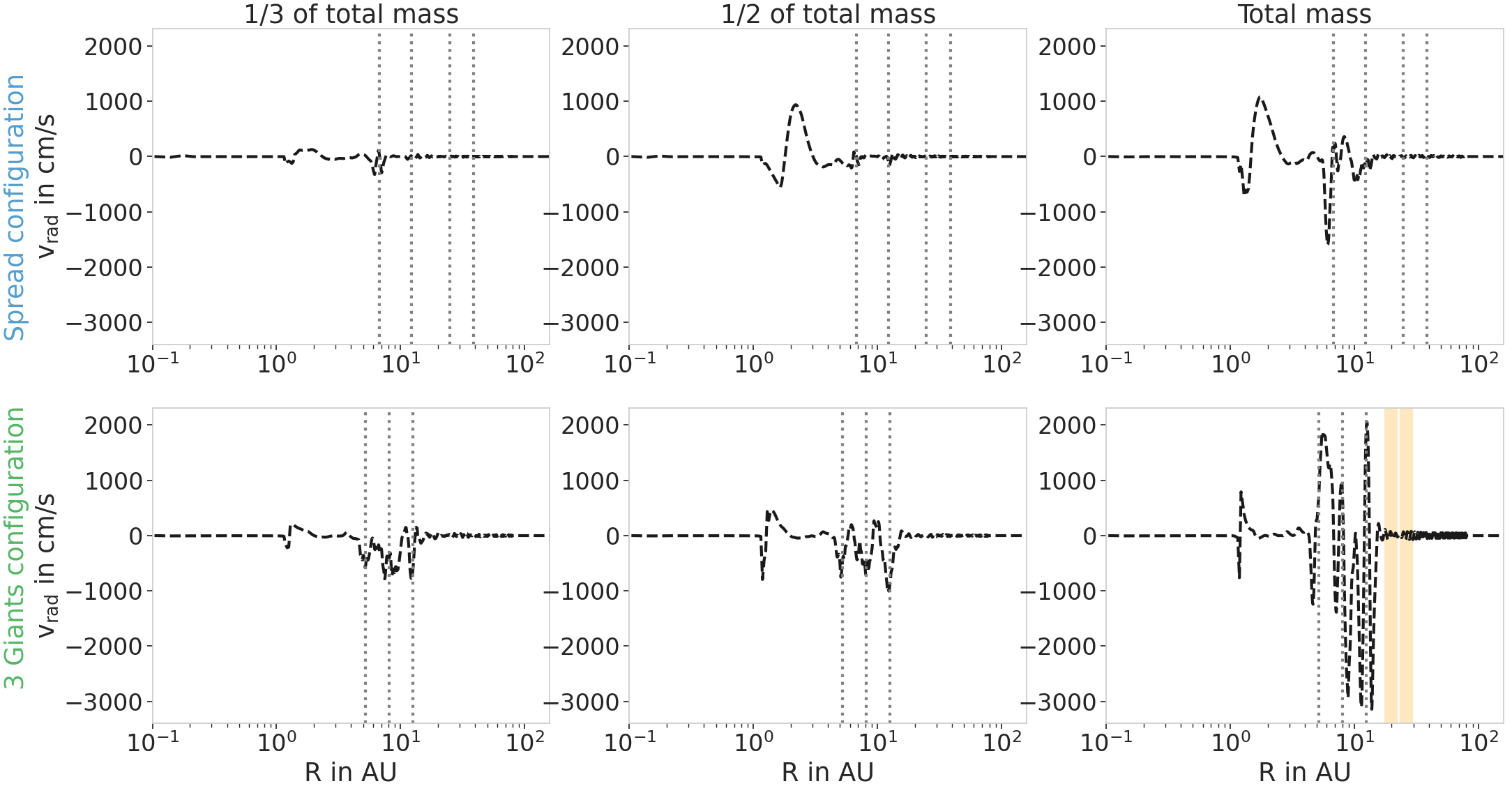}
   \caption{Same as Fig. \ref{vrad1D_hMMSNh4} but for the Spread and Three-Giants configurations and different planet masses. The two orange vertical lines show the positions of the rings observed in Fig. \ref{profile_diffmasses}.}
   \label{vrad1D_diffmass}
\end{figure*}

We show in Sect. \ref{section_dustevol} that the velocities of the planets can create traffic jams that produce noticeable substructures. This is due to the fact that the gas disk is highly perturbed by multiple giant planets. In Figs. \ref{vrad1D_hMMSNh4} and \ref{vrad1D_diffmass} we present the radial azimuthally and time-averaged profiles used in our simulations. We see that even after averaging the profiles for 2 500 orbits, the disk remain highly perturbed by the planets. In Fig. \ref{vrad1D_diffmass} we clearly show that these perturbations are due to the planets as we see that they are stronger for more massive planets.

\FloatBarrier

\begin{figure*}[t]
   \centering   
   \includegraphics[scale=0.31]{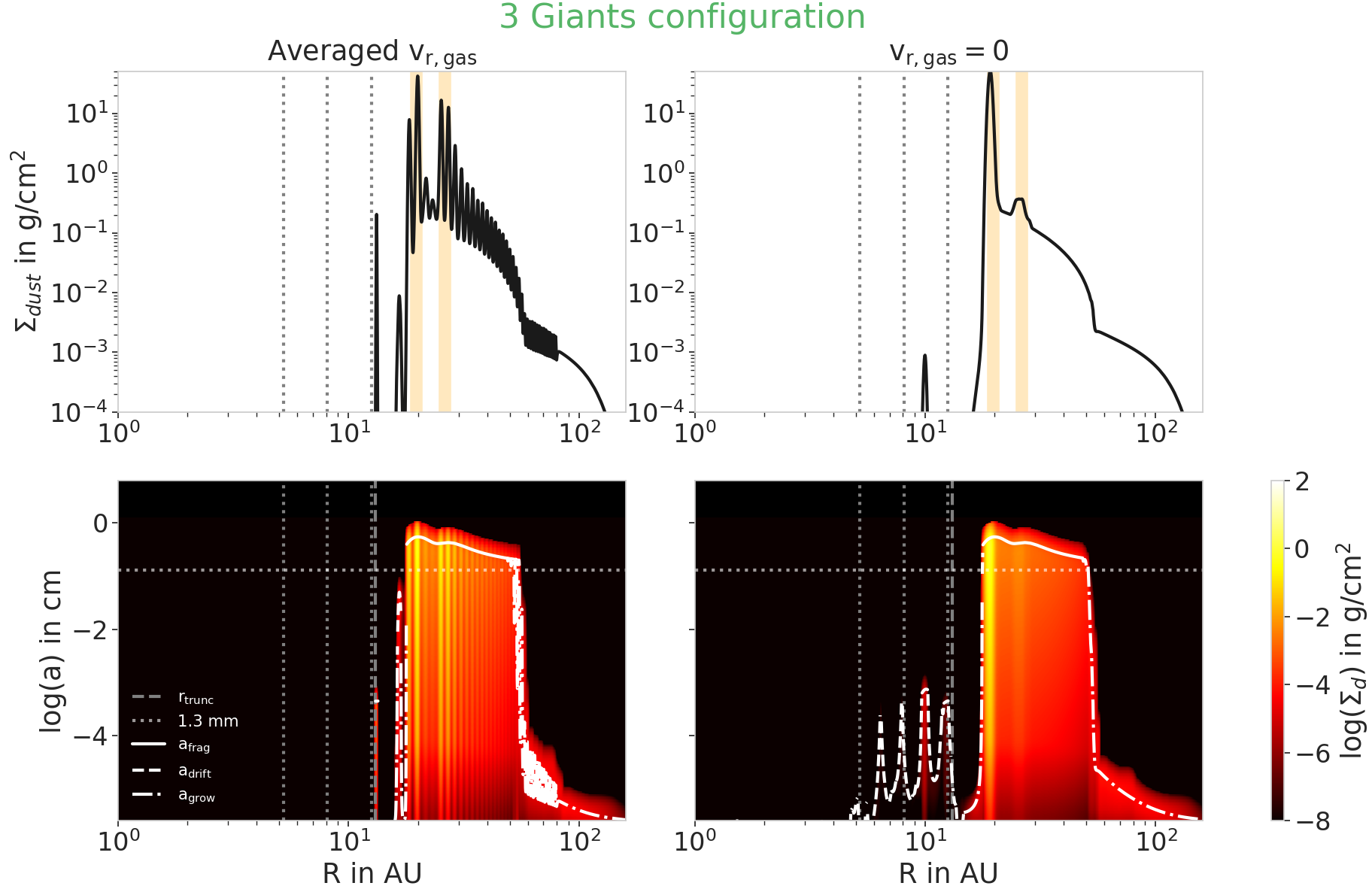}
   \caption{Dust distributions in the Three-Giants configuration case: in the left panels, the velocity of the gas used as an input for the dust evolution model is the averaged velocity as presented in Fig. \ref{vrad1D_diffmass}; in the right panels, the gas velocity is taken to be null. The first row presents the integrated dust distribution along all the dust sizes, representing the total dust distribution in the disk, whereas the second row presents the classic dust distributions as in Sect. \ref{section_dustevol}. The presence of spikes in the left panel shows that the radial gas velocities are indeed responsible for the dust accumulations in additional rings exterior to the positions of the planets.}
   \label{dist_vradVSzero}
\end{figure*}

\section{Impact of the radial gas velocity on the dust distributions}
\label{appendix_gasvrad}

In Sect. \ref{section_dustevol} some over-densities are observed at positions that are not directly related to the orbits of the planets or to any perturbations in the gas surface density. These over-densities originate in the radial velocity profile of the gas disk, highly perturbed by the presence of multiple giant planets. These perturbations create traffic jams, where the dust can accumulate without being trapped. In order to determine if these traffic jams are indeed producing such over-densities, we study the dust evolution distribution also with a gas radial velocity forced to be zero. 

We take the example of the Three-Giants configuration as it produces the most perturbed disk. We present in Fig. \ref{dist_vradVSzero} the dust distributions in the case where the same gas surface density profile is given to the model but the radial velocity profiles are either averaged as in this paper (left panels) or set to zero (right panels). In the first row, we show the integrated dust surface density over all the grain sizes: these profiles allow us to see that the dust is distributed differently in both cases. When the radial gas profile is set to zero, the dust mostly accumulate in the pressure bumps present in the disk. Even if the gas surface density profile present a very slight bump located at 26 AU, creating a small over-density in the dust at this location, it is too small to create a noticeable feature in the observations. However, when the gas radial profile is taken into account, the dust gets stuck in these different traffic jams, explaining this spiky behavior. When compared to the positions of the rings observed specially in Fig. \ref{image_beams}, represented in orange in this figure, we see that the second ring located at 26 AU is clearly originating in the strong traffic jam located at the same semimajor axis.

As these traffic jams can create noticeable substructures, we conclude that the gas radial velocity profile has a non-negligible impact on the dust distributions when multiple planets are present in the disk. This is important for the derivation of synthetic images but also for dust evolution models.

\FloatBarrier

\section{Complementary images}
\label{appendix_comp_images}

\subsection{Solar System images}
\label{appendix_images}

We show in this appendix the images of the Solar System images corresponding to the normalized intensity profiles presented in Sect. \ref{section_images}.

\begin{figure*}[h]
   \centering   
   \includegraphics[scale=0.33]{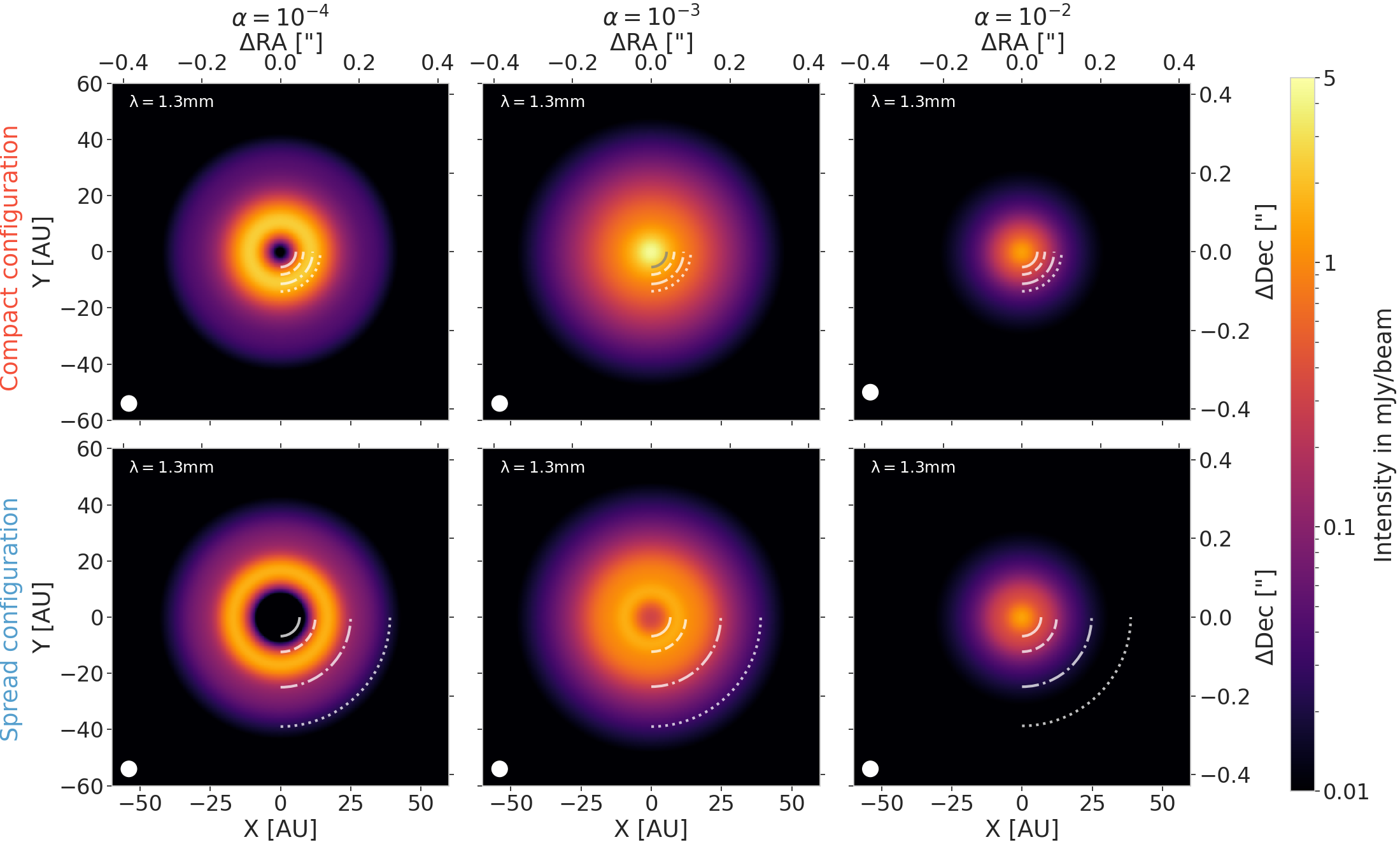}
   \caption{Images at $\lambda = \rm 1.3mm$ for each Solar System configuration for an MMSN-like aspect ratio. These are the images that correspond to the intensity profiles presented in Fig. \ref{profile_hMMSN}. The beam is 0.04"$\times$0.04" and is represented as the white circle in the lower-left corner of each panel. The white lines represent the distances of the different planets.}
   \label{image_1.3mm_hMMSN}
\end{figure*}

\begin{figure*}[h]
   \centering   
   \includegraphics[scale=0.33]{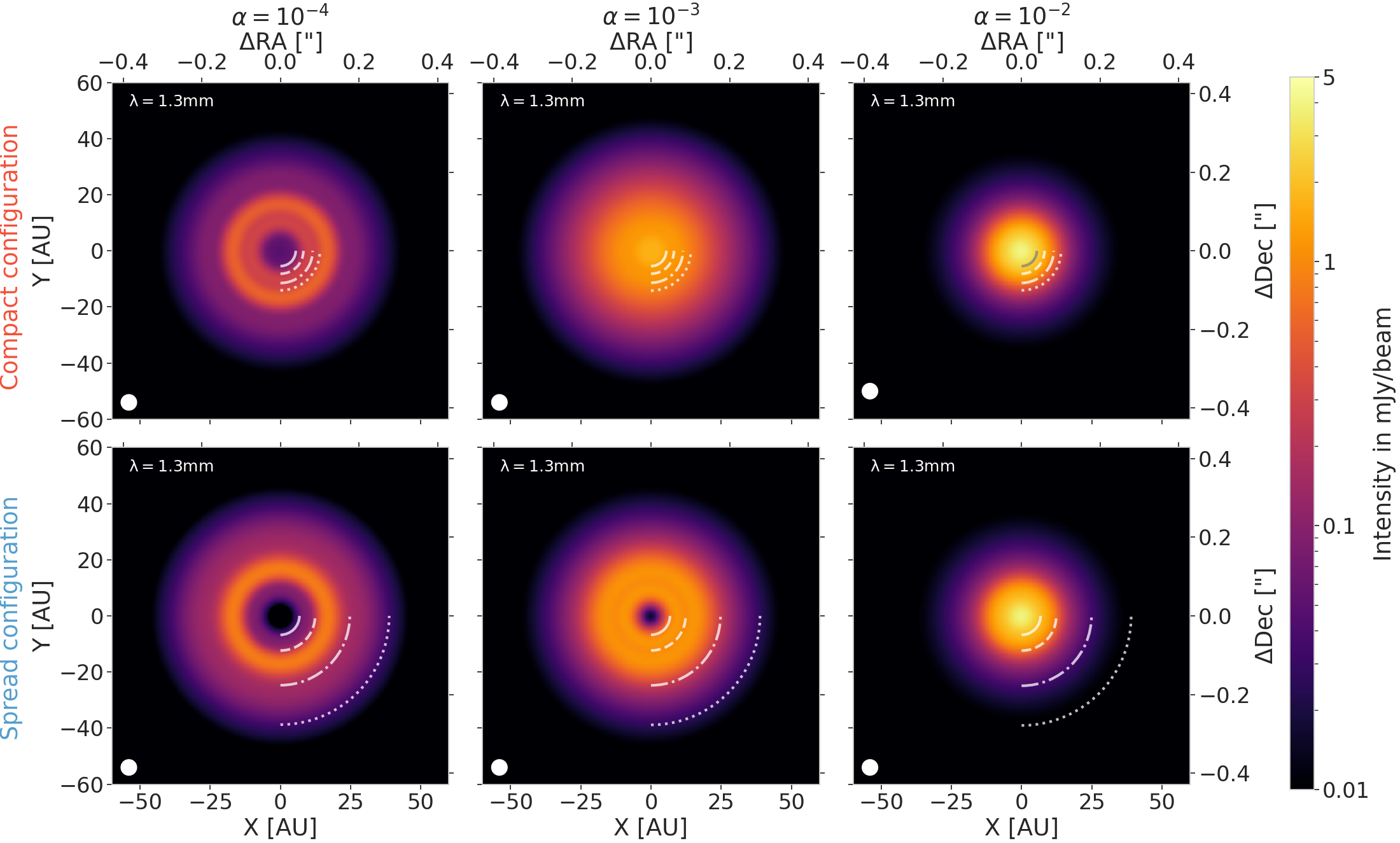}
   \caption{Same as Fig. \ref{image_1.3mm_hMMSN} but for a smaller aspect ratio.}
   \label{image_1.3mm_h4}
\end{figure*}

\subsection{Images of inclined disks}
\label{appendix_images_inclined}

Here, we present the images of the disks with different inclinations in the Spread and Three-Giants configurations. These images correspond to the radial profiles presented in Fig. \ref{profile_inclinations}.

\begin{figure*}[h]
   \centering   
   \includegraphics[scale=0.265]{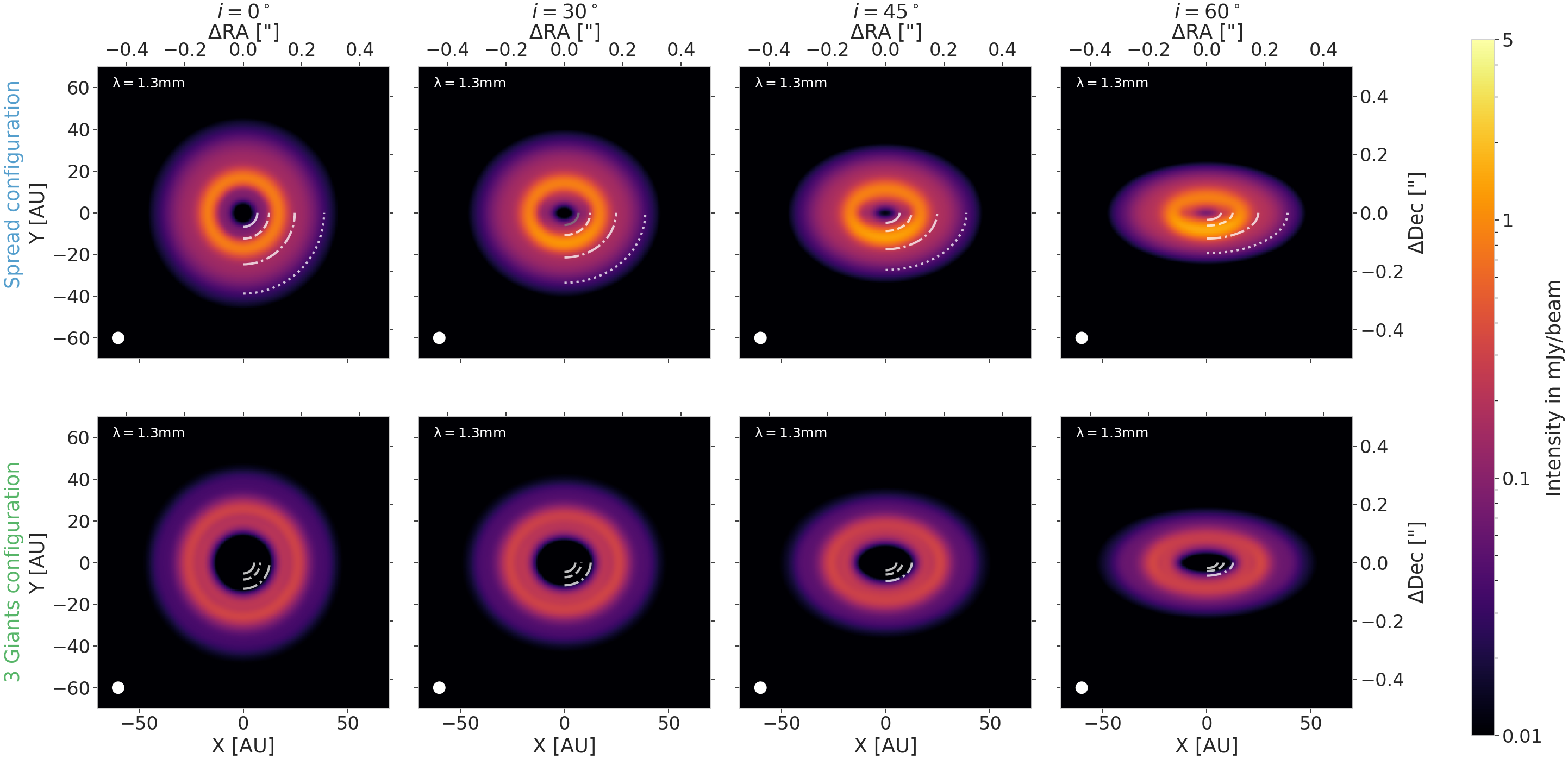}
   \caption{Images at $\lambda =$1.3mm in the Spread (first row) and Three-Giants (second row) configuration, at low viscosity and low aspect ratio, for different inclinations. The inclination is increasing from left to right, going from a face-on disk ($i = 0^\circ$) to a highly inclined disk ($i = 60^\circ$). The white lines represent the positions of the different planets in each configuration. }
   \label{image_differentinclinations}
\end{figure*}

\end{appendix}


\end{document}